\documentclass[aps, prd, reprint, onecolumn, nofootinbib]{revtex4-2}
\usepackage{amsmath}
\usepackage{amsfonts}
\usepackage{amssymb}
\usepackage{graphicx}
\usepackage{mathrsfs}
\usepackage{slashed}
\usepackage{xcolor}
\usepackage{subcaption}
\usepackage{relsize}
\newcommand{\braket}[1]{\Big<#1\Big>}
\newcommand{\Braket}[1]{\Big<#1\Big>}

\newcommand{\e}{\textrm{e}}

\newcommand{\symb}{\textrm{symb}^{-1}}
\newcommand{\non}{\nonumber\\}

\newcommand{\dif}{\mathlarger{\blacktriangle}}
\newcommand{\diff}{\mathlarger{\mathlarger{\blacktriangle}}}

\def\bear{\begin{eqnarray}}
\def\ear{\end{eqnarray}}
\def\e{{\rm e}}

\newcommand{\ds}{\slashed{\partial}}
\newcommand{\dsp}{\ds^{\prime}}
\newcommand{\Ds}{\slashed{D}}
\newcommand{\Dsp}{\Ds^{\prime}}
\newcommand{\Dspg}{\slashed{D}^{\gamma \prime}}
\newcommand{\As}{\slashed{A}}
\newcommand{\Js}{\slashed{J}}

\newcommand{\kb}{\bar{k}}

\newcommand{\be}{\begin{equation}}
\newcommand{\ee}{\end{equation}}
\newcommand{\bi}{\begin{itemize}}
\newcommand{\ei}{\end{itemize}}
\newcommand{\bea}{\begin{eqnarray}}
\newcommand{\eea}{\end{eqnarray}}

\begin{document}

\title{The Generalised LKF Transformations for Arbitrary $N$-point Fermion Correlators}

\author{Naser  Ahmadiniaz}
\email{n.ahmadiniaz@hzdr.de}
\affiliation{Helmholtz-Zentrum Dresden-Rossendorf, Bautzner Landstra\ss e 400, 01328 Dresden, Germany}
\author{James P. Edwards}
\email{jedwards@ifm.umich.mx}
\affiliation{Instituto de F\'isica y Matem\'aticas
Universidad Michoacana de San Nicol\'as de Hidalgo
Edificio C-3, Apdo. Postal 2-82
C.P. 58040, Morelia, Michoac\'an, M\'exico}
\author{José Nicasio}
\email{jose.nicasio@umich.mx}
\affiliation{Instituto de F\'isica y Matem\'aticas
Universidad Michoacana de San Nicol\'as de Hidalgo
Edificio C-3, Apdo. Postal 2-82
C.P. 58040, Morelia, Michoac\'an, M\'exico}
\author{Christian Schubert}
\email{schubert@ifm.umich.mx}
\affiliation{Instituto de F\'isica y Matem\'aticas
Universidad Michoacana de San Nicol\'as de Hidalgo
Edificio C-3, Apdo. Postal 2-82
C.P. 58040, Morelia, Michoac\'an, M\'exico}



\begin{abstract}
We examine the non-perturbative gauge dependence of arbitrary configuration space fermion correlators in quantum electrodynamics (QED). First, we study the dressed electron propagator (allowing for emission or absorption of any number of photons along a fermion line) using the first quantised approach to quantum field theory and analyse its gauge transformation properties induced by virtual photon exchange. This is then extended to the $N$-point functions where we derive an exact, generalised version of the fully non-perturbative Landau-Khalatnikov-Fradkin (LKF) transformation for these correlators. We discuss some general aspects of application in perturbation theory and investigate the structure of the LKF factor about $D = 2$ dimensions.
\end{abstract}

\maketitle

\section{Introduction}
The non-perturbative structure of the $N$-point functions in QED is an important aspect of quantum field theory, yet analysing such aspects of the theory remains a difficult problem and still attracts significant attention. In a general theory, such information plays an important role in determining its phase structure, such as for dynamical chiral symmetry breaking or confinement in the well-known example of QCD. It is often desirable to determine the gauge dependence of various quantities, or to use results found in a certain gauge to extract information about the same quantity in a different gauge. There exist limited non-perturbative analyses of relatively simple objects such as the Ball-Chiu decomposition of the QED vertex \cite{BallChiu} and its perturbative determination at one-loop order in various gauges \cite{YFVertex, KizVertex}; results of similar calculations of the $3$-point vertex in three-dimensional QED$_{3}$ have been reported in \cite{QED31, QED33,QED34, QED35, aitchison1997inverse,mitra2006gauge} and for scalar QED in \cite{ScalarVertex1, ScalarVertex2}. Ball and Chiu generalised their work to analyse the one-loop quark gluon vertex in QCD \cite{BallChiuQCD1}, later extended to two-loop order \cite{VertexQCD1, VertexQCD2} in a particular renormalisation scheme and its gauge structure in arbitrary covariant gauge and dimension for an $SU(N)$ symmetry group \cite{Davydychev} -- this information is crucial for the determination of the transverse part of the vertex. Similarly it has been possible to calculate the three- and four-gluon vertices, off-shell, in covariant gauges for special kinematics up to two-loop order \cite{GluonVertex2,  GluonVertex4, GluonVertex5, GluonVertex6, GluonVertex7, GluonVertex10, GluonVertex11, GluonVertex12} which bears strongly on both the Dyson-Schwinger equations and infrared divergences within QCD.

Although physical observables such as cross sections are gauge invariant, the $N$-point Green functions of a given gauge theory generally have a strong dependence on the gauge choice for internal photons (the gauge transformations of external photons are well understood via the Ward-Takahashi \cite{WTI} or Slavanov-Taylor \cite{STI1, STI2} identities whereas perturbation theory requires the gauge of internal photons to be fixed in order to define their propagator). Various covariant gauges can offer significant advantages for specific computations: Feynman gauge, besides minimising the number of terms in loop calculations, leads to simple ghost-free Ward identities \cite{Binger}. Landau gauge has become the favourite in Yang-Mills theory and QCD since it leads to an UV-finite ghost-gluon vertex \cite{STI1} and an IR fixed point of the renormalisation group flow accessible on the lattice \cite{Lattice1, Lattice2}. The Yennie-Fried gauge ($\xi = 3$) \cite{YFG} is useful for eliminating spurious IR divergences in $D=4$ \cite{YFVertex} and a similar role is played by the traceless gauge $\xi = -1$ in $D=2$ \cite{WLEHL}. It is thus of great interest to develop efficient techniques for transforming Green's functions from one gauge to another. 
Such studies were initiated for the QED $2$-point function -- or propagator --  and the electron-photon vertex for the family of linear covariant gauges by Landau and Khalatnikov \cite{LandauLKFT} and independently by Fradkin \cite{FradkinLKFT} and were later revisited using functional techniques in \cite{ZuminoLKFT1, ZuminoLKFT2} -- see also \cite{juer1983zumino}.

These ``LKF'' transformations are framed in coordinate space and are fully non-perturbative: denoting the propagator with covariant gauge parameter $\xi$ by $S(x; \xi)$ a variation in the gauge $\xi \rightarrow \xi + \Delta \xi$ leads to the transformation ($\xi = 0$ corresponds to Landau gauge and $\xi = 1$ to Feynman gauge)
\begin{equation}
	S(x; \xi + \Delta \xi) = S(x; \xi) \e^{-i\Delta \xi\left[\Delta_{D}(x ) - \Delta_{D}(0)\right]}\, ,
	\label{eqLKFTProp}
\end{equation} 
where $\Delta_{D}(x)$ is a function that fixes the gauge \cite{LandauLKFT} (see (\ref{eqDeltaD})). Aside from specifying the way in which the $N$-point functions vary under a gauge transformation at a given loop order, the linear dependence of $\Delta_{D}$ on $\alpha$ shows that one can construct gauge dependent parts of higher-loop diagrams from knowledge of lower order terms \cite{AdnanProc, AdnanRaya}. The utility of these transformations is well illustrated by the restrictions imposed on the non-perturbative three-point vertex for it to be compatible with the LKF transformation of the fermion propagator \cite{Curtis, Dong, Kizilersu}. Unfortunately, despite their importance in restraining the structure of gauge theory interactions, the LKF transformations have been studied far less than the more familiar constraints arising from the Ward-Takahashi / Slavanov-Taylor identities \cite{Aguilar, Gracey}. 

However, the LKF transformations for the particular case of the propagator have been examined to varying degrees of detail. In the massless case, it has been shown for both scalar and spinor QED and in particular circumstances for QCD that wavefunction renormalisation takes a multiplicative power law form in four dimensions \cite{AdnanProc, Aslam, Bloch}. Moreover, the LKF transformations link the wave function normalisation constants in different gauges \cite{ZuminoLKFT1} and relate strongly to chiral symmetry breaking. It has also been shown that the LKF rules for the fermion propagator lead to an enhancement to the quark anomalous magnetic moment \cite{Chang}. Results for the propagator in reduced QED are given in \cite{RQED1, QED35}. 

Going beyond the propagator, there has been a recent resurgence in studying the generalised LKF transformations for $N$-point functions, especially in the context of QCD and extensions of the Gribov-Zwanziger (GZ) scenario \cite{gribov1978quantization,zwanziger1989local, dudal2008refinement} away from Landau gauge \cite{Gribov1, Gribov2} amongst other non-perturbative properties of Yang-Mills theories.  The use of the LKF transformations in perturbation theory is discussed in detail in \cite{Kissler, KissDirk}. More modern treatments have arrived at the transformation rules by including auxiliary ``Stueckelberg-type'' fields and BRST invariance \cite{Sonoda, Pietro1} (see also \cite{Ply1, Ply2}), which methods were recently applied to the gluon propagator \cite{Pietro2}. 

The LKF transformations in scalar QED for both the propagator and the generalised case of the $N$-point functions were also derived in \cite{LKFTWL1, LKFTWL2} using the alternative \textit{worldline approach} to quantum field theory. There it is shown that the ordered (quenched) $(N = 2n)$-point amplitudes completely fix the LFK transformations. Denoting such an amplitude that corresponds to the contraction of $n$ fields $\phi(x_{i})$ with $n$ conjugate fields $\phi^{\dagger}(x^{\prime}_{\pi(i)})$ for $\pi \in S_{n}$ as $\mathcal{A}(x_{1}, \ldots , x_{n}; x^{\prime}_{\pi(1)},\ldots, x^{\prime}_{\pi(n)} | \xi)$ when covariant gauge parameter $\xi$ is used for internal photons they find the amplitudes in different gauges are related by
\begin{equation}
	\mathcal{A}(x_{1}, \ldots , x_{n}; x^{\prime}_{\pi(1)},\ldots, x^{\prime}_{\pi(n)} | \xi + \Delta \xi) = \prod_{k, l = 1}^{N}\e^{-\Delta_{\xi}S_{i \pi}^{(k, l)}} \mathcal{A}(x_{1}, \ldots , x_{n}; x^{\prime}_{\pi(1)},\ldots, x^{\prime}_{\pi(n)} | \xi )
	\label{eqScalarLKFT}
\end{equation}  
which is the natural generalisation of (\ref{eqLKFTProp}), constructed now from functions $\Delta_{\xi}S_{i \pi}^{(k, l)}$ to be defined below in (\ref{eqDeltaS}). This same worldline formalism was recently applied to extend this work to the case of spinor QED \cite{UsLKFT1}. In this companion paper we provide further calculational details on this worldline derivation of the LKF transformation of the fermion propagator and the $N$-point fermionic Green functions that enter the calculation of scattering amplitudes in perturbation theory and discuss various applications of these recent results. 

The worldline formalism is an alternative, first quantised approach to quantum field theory that has its roots in work due to Feynman at the same time that the more familiar second quantised approach was developed \cite{FeynSc, FeynSp}. Strassler later developed perturbation theory within this framework \cite{Strassler1}, motivated by the seminal works of Bern and Kosower \cite{BK1, BK2}. The essential idea is to re-express the field theory scattering matrix in terms of path integrals over relativistic point particle trajectories which reproduce the so-called Master Formulae of Bern and Kosower -- a detailed review and a more recent report describing these methods can be found in \cite{ChrisRev, UsRev}. Despite remaining lesser known than the ``standard'' perturbation theory based on Feynman diagrams that came to dominate the development of quantum field theory, the first quantised worldline approach has been applied with great success in a wide variety of problems. Of particular importance in the context of LKF transformations, the non-trivial task of extracting the form factor decomposition of the QCD vertex that is usually done by analysing the Ward identities is quite cumbersome for rich tensorial structures. The worldline calculation has shown its efficiency in the decomposition of the three- and four-gluon vertices and the generalisation of the $N$-point Ward identity -- see \cite{GluonVertex5, GluonVertex10,GluonVertex12} -- and we expect similar simplifications in analyses of the gauge structure of the propagator. 

Early successful applications of the worldline approach include a semi-classical ``instanton'' based determination of the Schwinger pair production rate \cite{AAM}. By now it is recognised that the worldline formalism has several advantages over standard methods, including representations of amplitudes where virtual momenta have already been integrated over that combine multiple Feynman diagrams related by permutation of external legs \cite{Daikouji, Schmidt, WLEHL, UsRev, WLNotes} and are gauge invariant even at the level of the integrand \cite{Strassler1, ChrisRev}. Initial work using worldline techniques was largely focussed on loop amplitudes and although there are a few preliminary representations of propagators and tree-level processes using worldline techniques \cite{Daikouji, Sherry, vanHolten, Alexandrou, FradGitSymb, Fainberg, Tsai}, it is only recently that a complete description of the scalar \cite{Ladder,  LKFTWL1, LKFTWL2, Aftab, ScalarTadpole, Confined, Seba}, spinor \cite{fppaper1, fppaper2, SpinorTadpole, Reducible} and quark \cite{Karanikas2, ColTree} propagators that are needed to study the LKF transformations has been achieved that retains the familiar benefits of the first quantised approach (see also \cite{Somdatta}). 

We are motivated by several main goals. Firstly this article expands upon the brief report of the main results given in \cite{UsLKFT1} where we sought to compare the forms of the generalised transformation of the $N$-point functions between spinor and scalar QED; secondly the theoretical developments presented here for spinor QED are far from trivial and will serve as a stepping stone to the more complicated transformations in QCD or more general gauge theories (worldline techniques have been extended to the non-Abelian case in a series of recent articles \cite{Col1, Col2, JO1, JO2, UN, Mueller2}); finally a systematic study of the variation of $N$-point functions under a change of gauge is crucial for understanding how gauge invariant information can be extracted from calculations carried out in particular gauges -- we have in mind, for instance, the truncation of Dyson-Schwinger equations to a particular order or numerical evaluation of such quantities on the lattice. Our use of the worldline formalism will be seen to simplify both the derivation of the LFK transformations and their implementation in perturbation theory. 

\subsection{Overview}
The thrust of our approach and the main results can be summarised as follows. The worldline representation of the spinor propagator was recently developed in \cite{fppaper1, fppaper2} and uses the second order formalism of the Dirac field \cite{SO1, SO2}. The fundamental procedure is to decompose the configuration space representation of a Dirac fermion propagating in an electromagnetic field with gauge potential $\textrm{A}(x) = A_{\mu}(x)dx^{\mu}$ as
\begin{equation}
	S^{x'x}[A] = \big[m + i\Dsp\big]K^{x'x}[A]\, ,
	\label{eqSProp}
\end{equation}
where $D^{\prime}_{\mu}(x) = \partial^{\prime}_{\mu} + ieA_{\mu}(x')$ is the covariant derivative acting at $x^{\prime}$ and $K^{x'x}[A]$ is a matrix-valued auxiliary kernel to be discussed below. Here we shall describe $N$-point functions by extending this representation to multiple open fermionic lines. To isolate the gauge transformation of internal photons it is convenient to utilise the background field method \cite{Abbott1, Abbott2} to split the gauge field $A = \textrm{A}^{\gamma} + \bar{\textrm{A}}$ into a part $\textrm{A}^{\gamma}$ representing external photons and a ``quantum'' piece $\bar{\textrm{A}}$ -- the path integral over $\bar{\textrm{A}}$ will produce the virtual photons joining the spinor lines. These photons' propagators will be evaluated in a particular covariant gauge with parameter $\xi$ and we examine the dependence of the propagator on this choice. Related to this we define the ``backgroundless'' propagator by 
\begin{equation}
	S_{0}^{x'x}[A^{\gamma} + \bar{A} ; \xi] = \big\langle \big[m + i\big(\ds^{\prime} + ie\As^{\gamma}\big)\big]K^{x'x}[A^{\gamma} + \bar{A}] \big\rangle_{\bar{A},\, \xi}\, , 
\end{equation}
without the insertion of $-e\bar{\As}(x')$ in the prefactor of (\ref{eqSProp}). The expectation value is taken over configurations of the background field that produces loop photons attached to the fermion line with the photon propagator taken in the covariant gauge with parameter $\xi$. 

Our principle results are the following. The backgroundless part of the $(N = 2n)$-point function will be shown to transform according to the following rules for its partial amplitudes (again $\pi \in S_{n}$)
\begin{align}
	&\mathcal{S}_{0\pi}\left(x_{1}, \ldots , x_{n}; x^{\prime}_{\pi(1)},\ldots, x^{\prime}_{\pi(n)} |  \xi + \Delta \xi \right) \nonumber \\
	 = &\Big< \prod_{i = 1}^{n}\big[m + i\big(\ds^{\prime}_{i} + ie \As^{\gamma}\big)\big]K^{x^{\prime}_{\pi(i)} x_{i}}[A^{\gamma} + \bar{A}]\Big>_{\bar{A}, \,\xi + \Delta \xi} \nonumber \\
	  =&\Big< \prod_{i = 1}^{n}\big[m + i\big(\ds^{\prime}_{i} + ie \As^{\gamma}\big)\big]K^{x^{\prime}_{\pi(i)} x_{i}}[A^{\gamma} + \bar{A} ]\Big>_{\bar{A}, \,\xi} \prod_{k, l = 1}^{N}\e^{-\Delta_{\xi}S_{i \pi}^{(k, l)}} 
	\label{eqS0LKFT}
\end{align}
with the \textit{same} scalar factor as in (\ref{eqScalarLKFT}). In this case, however, the derivatives $\ds^{\prime}_{i} := \gamma^{\mu}\partial_{x_{i}^{\prime \mu}}$ act through onto the trailing exponential and generate additional terms involving the derivatives of the $\Delta_{\xi}S_{i \pi}^{(k, l)}$. The additional contractions that appear from the $\bar{\As}$ multiplying the kernels $K^{ x_{\pi(i)} x_{i}}$ in the complete $\mathcal{S}^{ x_{\pi(i)} x_{i}}$ precisely cancel these to allow for the exponential factor to be commuted to the left of the expectation value in (\ref{eqS0LKFT}). In this way we arrive at the generalised LKF transformations of the partial fermionic $N$-point functions,
\begin{align}
	&\mathcal{S}_{\pi}\left(x_{1}, \ldots , x_{n}; x^{\prime}_{\pi(1)},\ldots, x^{\prime}_{\pi(n)} | \xi + \Delta \xi  \right) \nonumber \\ 	
	= &\Big< \prod_{i = 1}^{n}\big[m + i\big(\ds^{\prime}_{i} + ie (\As^{\gamma} + \bar{\As} )\big)\big]K^{x^{\prime}_{\pi(i)} x_{i}}[A^{\gamma} + \bar{A}]\Big>_{\bar{A}, \, \xi + \Delta \xi} \nonumber \\
	 =& \prod_{k, l = 1}^{N}\e^{-\Delta_{\xi}S_{i \pi}^{(k, l)}} \Big< \prod_{i = 1}^{n}\big[m + i\big(\ds^{\prime}_{i} + ie (\As^{\gamma} + \bar{\As})\big)\big]K^{x^{\prime}_{\pi(i)}x_{i}}[A^{\gamma} + \bar{A} ]\Big>_{\bar{A},\, \xi}\nonumber \\
	  =&\prod_{k, l = 1}^{N}\e^{-\Delta_{\xi}S_{i \pi}^{(k, l)}} \mathcal{S}_{\pi}\left(x_{1}, \ldots , x_{n}; x^{\prime}_{\pi(1)},\ldots, x^{\prime}_{\pi(n)} | \xi  \right)
	\label{eqSLKFT}
\end{align}
which is the direct generalisation of (\ref{eqScalarLKFT}) to the spinor case reported in \cite{UsLKFT1} (we emphasise that the exponential factor is identical to the scalar result). 

Moreover, we shall show that the exponential prefactor, once summed over $k$ and $l$, is independent of the permutation, and thus factorises out of the sum over partial amplitudes, giving a simple multiplicative transformation for the $N$-point correlator itself, denoted $S$,
\begin{equation}
	S\left(x_{1}, \ldots , x_{n}; x^{\prime}_{1},\ldots, x^{\prime}_{n} | \xi + \Delta \xi  \right) = T_{n}S\left(x_{1}, \ldots , x_{n}; x^{\prime}_{1},\ldots, x^{\prime}_{n} | \xi \right) \, ,
	\label{eqSLKFTGlobal}
\end{equation}
with $T_{n}$ the exponential prefactor for any chosen permutation. Likewise, a similar simplification lifts the result for the partial amplitudes of scalar QED, (\ref{eqScalarLKFT}), to the complete scalar propagator. Indeed the congruence of these result mirrors the original outcome of Landau and Khalatnikov's analysis \cite{LandauLKFT} which makes clear that the propagator's transformation is essentially independent of the particular field theory under study; here we prove this to hold for arbitrary correlators. 

This paper has the following structure: in section \ref{secLKFT} we review the precise form of the LKF transformations and recent work on their application. In section \ref{secFProp} we use the recently developed worldline representation of the fermion propagator \cite{fppaper1} to study its gauge transformation properties, followed by the generalisation to the LKF transformations of the $N$-point functions in section \ref{secNPoint}. We then illustrate the application of our results in perturbation theory in section \ref{secPert} before a conclusion and discussion of ongoing and future work. 

\section{Gauge transformations of Green functions}
\label{secLKFT}
The LKF transformations show how field theory Green functions change between linear covariant gauges and contain information about their gauge-dependent part to all orders in the coupling to internal gauge bosons. We begin by reviewing the original construction of the coordinate space LKF transformations for the propagator ($2$-point functions) presented in \cite{LandauLKFT, FradkinLKFT, ZuminoLKFT2}. 

The coordinate space photon propagator corresponding to the covariant gauge parameter $\xi$ can be written
\begin{equation}
	G_{\mu\nu}(x - x^{\prime}; \xi) := \langle \bar{A}_{\mu}(x)\bar{A}_{\nu}(x^{\prime})\rangle_{\xi} = G_{\mu\nu}(x - x^{\prime} ; \hat{\xi}) + \Delta \xi \partial_{\mu}\partial_{\nu}\Delta_{D}(x - x^{\prime})\, ,
	\label{eqGPhoton}
\end{equation}
where $\hat{\xi}$ refers to a reference covariant gauge (chosen arbitrarily) and $\Delta \xi = \xi - \hat{\xi}$. Here $\Delta_{D}$ is a function that fixes the gauge (see \cite{LandauLKFT, ValeryDelta}), given by 
\begin{equation}
	\Delta_{D}(y) = -ie^{2}(\mu)\mu^{4-D}\int d^{D}\kb\,  \frac{\e^{-i k \cdot y}}{k^{4}} = -\frac{ie^{2}(\mu)}{16\pi^{\frac{D}{2}}}\Gamma\Big[\frac{D}{2} - 2\Big](\mu y)^{4-D}\, ,
	\label{eqDeltaD}
\end{equation}
where we used $ d^{D}k = (2\pi)^{D}d^{D}\kb$ for brevity and introduced the mass scale $\mu$ by identifying $e^{2} \rightarrow \mu^{4-D}e^{2}(\mu)$ to maintain a dimensionless coupling constant $e(\mu)$. The relation (\ref{eqGPhoton}) can be found by considering a gauge transformation $\bar{A}_{\mu} \rightarrow \bar{A}_{\mu} - \partial_{\mu}\phi$ and interpreting the function $\phi$ as a Stueckelberg-type scalar field. Quantisation in momentum space in covariant gauge with parameter $\xi$ yields \cite{Pietro1, CapriPhi} the correlation functions
\begin{align}
	\langle \bar{A}_{\mu}(k)\phi(-k)\rangle_{\xi} &= \frac{i\xi}{k^{4}} k_{\mu}\\
	\langle \phi(k)\phi(-k)\rangle_{\xi} &= \frac{\xi}{k^{4}} 
\end{align}
so that under $\bar{A}_{\mu}(k) \rightarrow \bar{A}_{\mu}(k) -ik_{\mu}\phi(k)$ the photon two-point function changes according to
\begin{equation}
	\langle \bar{A}_{\mu}(k)\bar{A}_{\nu}(-k)\rangle_{\xi} \longrightarrow  \langle \bar{A}_{\mu}(k)\bar{A}_{\nu}(-k)\rangle_{\xi} - \xi \frac{k_{\mu}k_{\nu}}{k^{4}}\, .
\end{equation}
This reproduces the Fourier space representation of $\partial_{\mu}\partial_{\nu}\Delta_{D}$ with the familiar result that it is only the longitudinal part\footnote{We refer to the usual momentum space decomposition of the photon propagator ${G_{\mu\nu}(k) = \Delta(k^{2})P_{\mu\nu} + \xi/k^{2} L_{\mu\nu}}$ into its transverse projector $P_{\mu\nu} := \delta_{\mu\nu} - \frac{k_{\mu}k_{\nu}}{k^{2}}$ and longitudinal part, $L_{\mu\nu}:= \frac{k_{\mu}k_{\nu}}{k^{2}}$ where at tree level, of course, $\Delta(k^{2}) = \frac{1}{k^{2}}$.} of the photon propagator that varies with $\xi$. 

It is the function $\Delta_{D}$ that appears in (\ref{eqLKFTProp}) that transforms the matter field's propagator between covariant gauges. For $D = 3$, $\Delta_{3}(x) = -\frac{i}{2}\alpha x$ with the familiar fine structure constant, $4\pi\alpha = e^{2}$, so that (\ref{eqGPhoton}) gives \cite{QED33}
\begin{equation}
	S(x; \xi) = S(x; \hat{\xi})\e^{-\frac{\Delta \xi}{2}\alpha x}.
	\label{eqSLKF3}
\end{equation}
In four dimensions, however, $\Delta_{D}$ requires regularisation; expanding (\ref{eqDeltaD}) about $D = 4-\epsilon$ leads to the result \cite{QED33} ${\Delta_{4}(x) = \frac{i\alpha}{4\pi}\left[\frac{2}{\epsilon} + \gamma_{E} + \ln(\pi) + 2\ln(\mu x) + \mathcal{O}(\epsilon)\right]}$ so that, upon introducing a cut-off to regularise the $x$-dependent logarithm we arrive firstly at
\begin{equation}
	\Delta_{4}(x_{0}) - \Delta_{4}(x) = -i\ln \left[\frac{x^{2}}{x^{2}_{0}}\right]^{\frac{\alpha}{4\pi}}\, ,
\end{equation}
and subsequently to the LKF transformation
\begin{equation}
	S(x; \xi) = S(x; \hat{\xi})\left[\frac{x^{2}}{x^{2}_{0}}\right]^{-\frac{\Delta \xi \alpha}{4\pi}}.
	\label{eqSLKF4}
\end{equation}
As is clear in the original derivations of the LKF transformations this transformation is the same for scalar and spinor QED. We shall shortly rederive this result using worldline techniques. 

\subsection{Transformation of $N$-point functions}
The LKF transformation of the $N$-point correlation functions has been studied to differing extents in scalar and spinor QED and QCD. Here we shall review the calculation in scalar QED presented in \cite{LKFTWL1, LKFTWL2} where a first quantised approach was used to derive the transformation for arbitrary correlators but we refer to \cite{Pietro1, Pietro2} and references therein for an examination of these transformations using BRST symmetry in the standard formulation. 

The position space worldline representation of the partial $N = 2n$-point function in which the field $\phi(x_{i})$ is connected to the conjugate field $\phi^{\dagger}(x'_{\pi(i)})$ for $i \in \{1, \ldots n\}$ and $\pi \in S_{n}$ is given in \cite{LKFTWL1, LKFTWL2} as
\begin{equation}
	\hspace{-1.5em}\mathcal{A}(x_{1}, \ldots , x_{n}; x^{\prime}_{\pi(1)},\ldots, x^{\prime}_{\pi(n)} | \xi 	) = \prod_{i = 1}^{n}\int_{0}^{\infty}\hspace{-0.25em}dT_{i}\, \e^{-m^{2}T_{i}} \hspace{-0.25em}\int_{x_{i}(0) = x'_{\pi(i)}}^{x_{i}(T_{i})= x_{i}}\hspace{-1.5em}\mathscr{D}x_{i}(\tau_{i})\, \e^{-\sum_{l = 1}^{n}\left(S_{0}^{l} + S_{\gamma}^{l}\right) - \sum_{k, l = 1}^{n}S_{i \pi}^{(k, l)}(\xi)}
	\label{eqAWL}
\end{equation}
where $m$ is the mass of the field and the path integral is over trajectories that travel between $x^{\prime}_{\pi(i)}$ and $x_{i}$ in (Schwinger) proper time $T$. The worldline action has been split up into the free particle actions
\begin{equation}
	S_{0}^{l}[x_{l}] = \int_{0}^{T_{l}}d\tau_{l} \, \frac{\dot{x}_{l}^{2}}{4},
	\label{eqS0}
\end{equation}
the interaction of these particles with external photons,
\begin{equation}
	S_{\gamma}^{l}[x_{l}, A^{\gamma}] = ie\int_{0}^{T_{l}}d\tau_{k} \, \dot{x}_{l}\cdot A^{\gamma}(x_{l}(\tau_{l}))\,,
	\label{eqAgam}
\end{equation}
and
\begin{equation}
	S_{i\pi}^{(k, l)}[x_{k}, x_{l}, \bar{A} | \xi] = \frac{e^{2}}{2}\int_{0}^{T_{k}}d\tau_{k}\int_{0}^{T_{l}}d\tau_{l}\, \dot{x}_{k}^{\mu}G_{\mu\nu}(x_{k} - x_{l}; \xi)\dot{x}_{l}^{\nu}\, ,
	\label{eqSi}
\end{equation}
which produces the electromagnetic interaction due to exchange of virtual photons between particle worldlines $k$ and $l$ in the chosen covariant gauge (we shall derive the equivalent action for spinor QED below). 

To study the LKF transformations we recall the explicit form of the configuration space photon propagator in an arbitrary covariant gauge,
\begin{equation}
	G_{\mu\nu}(y; \xi) = \frac{1}{(4\pi^{\frac{D}{2}})}\left\{ \frac{1 + \xi}{2}\Gamma\Big[\frac{D}{2} - 1\Big]\frac{\delta_{\mu\nu}}{y^{2}{}^{\frac{D}{2}-1}} + (1 - \xi)\Gamma\Big[\frac{D}{2}\Big] \frac{y_{\mu}y_{\nu}}{y^{2}{}^{\frac{D}{2}}}\right\}.
	\label{eqGConfig}
\end{equation}
Under a change in the gauge parameter, $\xi \rightarrow \xi + \Delta{\xi}$ the integrand of the action $S_{i}$ changes by a total derivative, ${S_{i\pi}^{(k, l)}(\xi) \rightarrow S_{i\pi}^{(k, l)}(\xi) + \Delta_{\xi}S_{i\pi}^{(k, l)}}$, where
\begin{align}
\hspace{-1.5em}	\Delta_{\xi}S_{i\pi}^{(k, l)} &= \frac{\Delta\xi e^{2}}{32 \pi^{\frac{D}{2}}} \Gamma\Big[\frac{D}{2} - 2\Big]\int_{0}^{T_{k}}d\tau_{k} \int_{0}^{T_{l}}d\tau_{l} \, \partial_{\tau_{k}}\partial_{\tau_{l}}\left[(x_{k}(\tau_{k}) - x_{l}(\tau_{l}))^{2}\right]^{2- \frac{D}{2}} \\
\hspace{-2em}		&= \frac{\Delta \xi e^{2}}{32 \pi^{\frac{D}{2}}} \Gamma\Big[\frac{D}{2} - 2\Big]\Big\{ \big[\left(x_{k} - x_{l}\right)^{2}\big]^{2-\frac{D}{2}} - \big[\left(x_{k} - x^{\prime}_{\pi(l)}\right)^{2}\big]^{2-\frac{D}{2}} \nonumber \\
	&\hspace{16.5em}- \big[\left(x^{\prime}_{\pi(k)} - x_{l}\right)^{2}\big]^{2-\frac{D}{2}} + \big[\left(x^{\prime}_{\pi(k)} - x^{\prime}_{\pi(l)}\right)^{2}\big]^{2-\frac{D}{2}}\Big\}\,.
	\label{eqDeltaS}
\end{align}
As we shall discuss below for the fermionic amplitudes, the effect of changing the gauge of the external photons also introduces a total derivative term in the action (see equation (\ref{eqDeltaV})). Consequently the contributions from gauge transforming the internal and external photons vanish for photons with at least one leg on a closed scalar loop. This allows us to focus on the quenched amplitudes, which completely fix the form of the LKF transformation, as discussed in \cite{LKFTWL1, LKFTWL2}. 

Even for quenched amplitudes there are further simplifications, since the gauge transformation of external photons is well understood through the Ward identity. As shall be made clear below, although such a gauge transformation produces a non-vanishing boundary contribution when the photon is attached to an open line, it does not contribute to the LSZ formula for on-shell matrix elements. As such, we may restrict our attention to the gauge transformations of only those virtual photons that mediate interactions between open lines. Then (\ref{eqDeltaS}), which does not depend upon the path integral variables nor the proper time, $T$, is the full contribution to the generalised LKF transformation,
\begin{equation}
	\mathcal{A}(x_{1}, \ldots , x_{n}; x^{\prime}_{\pi(1)},\ldots, x^{\prime}_{\pi(n)} | \xi + \Delta \xi	) = \prod_{k, l = 1}^{n}\e^{-\Delta_{\xi}S_{i\pi}^{(k, l)}}\mathcal{A}(x_{1}, \ldots , x_{n}; x^{\prime}_{\pi(1)},\ldots, x^{\prime}_{\pi(n)} | \xi 	) ,
	\label{eqScalarLKFTFinal}
\end{equation}
as reported in (\ref{eqScalarLKFT}); the full amplitude is constructed by simply summing the ordered amplitudes over the permutations $\pi \in S_{n}$. More information and a discussion about the pole structure in dimensional regularisation are given in \cite{LKFTWL1} and are elaborated for spinor QED below, where we will also compute the product over $k$ and $l$ explicitly. The application to perturbation theory in position space is also given in \cite{LKFTWL1} which we shall repeat for the spinor case presently.

\section{The fermion propagator in first quantisation}
\label{secFProp}
The fermion propagator has only recently been given a satisfactory worldline description that maintains the familiar advantages of the first quantised approach (see earlier attempts in \cite{vanHolten, Alexandrou, Fainberg}). Contrary to the one- or multi-loop case this involves a path integral over \textit{open} lines joining the endpoints of the propagation, and is also a function of the initial and final spin states. The path integral formulation of this construction is given in \cite{fppaper1, fppaper2} and the reader is referred to \cite{Olindofp} for an alternative approach.   

The fermion propagator in a background electromagnetic field $\textrm{A} = A_{\mu}dx^{\mu}$ is defined in position space by the matrix elements
\begin{equation}
	S_{\beta\, \alpha}^{x'x}[A] := \langle x', \beta | [m - i \Ds]^{-1} | x, \alpha \rangle\, ,
	\label{eqSMatrix}
\end{equation}
where the covariant derivative is given by $D_{\mu}:= \partial_{\mu} + i eA_{\mu}$ and $\alpha$ and $\beta$ indicate the spin at the points $x$ and $x'$ respectively. Applying the Gordon identity we can rewrite this as
\begin{align}
	S_{\beta \,\alpha}^{x'x}[A] &= \big[m+i\Dsp\big]_{\beta \sigma} \big\langle x', \sigma \big| \big[-\!\!D^{2} + m^{2}  + \frac{ie}{2} \gamma^{\mu}F_{\mu\nu}\gamma^{\nu} \big]^{-1} \big| x, \alpha \big\rangle \\
	&\equiv  \big[m+i\Dsp\big]_{\beta \sigma}K^{x'x}_{\sigma \alpha}[A]
	\label{eqSGordon}
\end{align}
where the covariant derivative acts on $x'$. The matrix element, which we call the kernel $K^{x'x}_{\sigma \alpha}$, now takes the form of the propagator for a scalar particle in the presence of a matrix valued potential and it is well known how to give a path integral representation for this object. As discussed in \cite{fppaper1} this path integral can be written as
\begin{equation}
	K^{x'x}[A]=2^{-\frac{D}{2}} \symb \int_{0}^{\infty} \!dT e^{-m^{2}T}  \!\int_{x(0) = x}^{x(T) = x'}\hspace{-1.5em}\mathscr{D}x(\tau)\int_{\psi(0) + \psi(T) = 0}\hspace{-3em}\mathscr{D}\psi(\tau) \,\e^{-\int_{0}^{T}d\tau\, \left[ \frac{\dot{x}^{2}}{4} + \frac{1}{2} \psi \cdot \dot{\psi} +ie \dot{x} \cdot A(x) -ie (\psi + \eta) \cdot F(x) \cdot (\psi + \eta)\right]}\,.
	\label{eqKPathInt}
\end{equation}
Here the path integrals are over trajectories from $x$ to $x'$, on which lives a one-dimensional field theory described by the action that couples the bosonic embedding coordinates $x^{\mu}(\tau)$ and the anti-periodic Grassmann variables $\psi^{\mu}(\tau)$ to the background field; the former produce the orbital interaction whilst the latter generate the spin coupling (the so-called Feynman ``spin factor'' \cite{FeynSp}). The spin structure of the kernel arises from the ``symbol map'' acting on the constant Grassmann variables $\eta^{\mu}$ according to
\begin{equation}
	\textrm{symb}\left\{ \gamma^{[\mu_{1}}\cdots \gamma^{\mu_{n}]} \right\} \equiv (-i\sqrt{2})^{n}\eta^{\mu_{1}}\ldots \eta^{\mu_{n}}\,,
\end{equation}
where the square brackets indicate anti-symmetrisation of the product with the appropriate combinatorial factor. As mentioned in \cite{fppaper1} this representation has the advantage of expressing the propagator directly in the Dirac basis of the Clifford algebra.

In the following we are interested in analysing scattering amplitudes involving an arbitrary number of external photons attached to the particle worldline and any number of virtual photon exchanges along the line. To achieve this we employ the background field method for the virtual photons, decomposing $\textrm{A} = \textrm{A}^{\gamma} + \bar{\textrm{A}}$ and we shall quantise $\bar{\textrm{A}}$ in the path integral formalism choosing a particular linear covariant gauge for the internal photons. The internal photons are thus produced by Wick contractions between distinct factors of $\bar{A}$ while external photons are represented by $A^{\gamma}$. Hence we write the full propagator (we subsequently suppress the spinor indices for brevity)
\begin{equation}
	S(x, x' | \xi) := \Big\langle S^{x'x}[A^{\gamma} + \bar{A}]\Big\rangle_{\bar{A},\, \xi} =  \Big\langle \big[m+i\Dspg - e\bar{\As}\big]K^{x'x}[A^{\gamma} + \bar{A}]\Big\rangle_{\bar{A},\, \xi}\, ,
	\label{eqSxi}
\end{equation}
where we have extracted the ``backgroundless'' part of the covariant derivative, ${D^{\gamma \prime}_{\mu} = \partial_{\mu}^{\prime} +ieA^{\gamma}_{\mu}}$. We follow the notation used to define (\ref{eqGPhoton}) and establish expectation values in the path integral approach in appendix \ref{AppExp}.

The final identification that can be made to connect this propagator to photon amplitudes is the specification of the external photon source as a sum of plane waves of fixed momenta, $k_{i\mu}$, with polarisations $\varepsilon_{i \mu}$,
\begin{equation}
	A_{\mu}^{\gamma}(x) = \sum_{i = 1}^{N} \varepsilon_{i\mu}\e^{i k_{i} \cdot x},
	\label{eqAPlane}
\end{equation}
after which the amplitude (\ref{eqSxi}) is expanded to multi-linear order in the polarisations. Substituted into $K[A^{\gamma} + \bar{A}]$ this leads to the insertion of photon vertex operators under the path integral (\ref{eqKPathInt}),
\begin{align}
	V^{x'x}_{\eta}[k, \varepsilon] &= \int_{0}^{T}d\tau \left[\varepsilon \cdot \dot{x}(\tau) - i(\psi(\tau) + \eta) \cdot f \cdot (\psi(\tau)+\eta)\right]\e^{i k \cdot x(\tau)} \\
	&= \int_{0}^{T}d\tau \, \e^{i k \cdot x(\tau) + \varepsilon \cdot \dot{x}(\tau) - i(\psi(\tau)+\eta) \cdot f \cdot (\psi(\tau)+\eta)}\big|_{\varepsilon}.
	\label{eqVertex}
\end{align}
Here we introduced the photon field strength tensor $f_{\mu\nu} := 2k_{[\mu}\varepsilon_{\nu]}$ and borrowed the trick often employed in string theory for such vertex operators by exponentiating the prefactor with the instruction that only the part linear in $\varepsilon$ should be taken. In this way (\ref{eqKPathInt}) becomes
\begin{align}
	\hspace{-2em}K^{x'x}[k_{1}, \varepsilon_{1}; \ldots ; k_{N}, \varepsilon_{N} | \bar{A}]&=(-ie)^{N} \int_{0}^{\infty} dT \e^{-m^{2}T} \int_{x(0) = x}^{x(T) = x'}\hspace{-1.5em}\mathscr{D}x(\tau) \,\e^{-\int_{0}^{T}d\tau\, \left[ \frac{\dot{x}^{2}}{4} +ie \dot{x} \cdot \bar{A}(x) \right]} \nonumber \\
	&\times \,2^{-\frac{D}{2}}\symb \int_{\psi(0) + \psi(T) = 0}\hspace{-3em}\mathscr{D}\psi(\tau)\, \e^{-\int_{0}^{T}d\tau \left[   \frac{1}{2} \psi \cdot \dot{\psi} -ie (\psi + \eta) \cdot \bar{F}(x) \cdot (\psi + \eta)\right]} \prod_{i=1}^{N}V_{\eta}^{x'x}[k_{i}, \varepsilon_{i}]\Big|_{\varepsilon_{1} \ldots \varepsilon_N}\,.
	\label{eqKPathIntV}
\end{align}
We have now separated the contributions to the kernel from external photons and the interaction with the background field, $\bar{\textrm{A}}$, that will produce the virtual photons running along the line. 
\subsection{Gauge transformations}
Here we consider how changing the gauge of the photons attached to the particle line affects the propagator.  To begin with, we consider the gauge transformation of the external photons represented by the vertex operators in (\ref{eqVertex}). The (momentum-space) gauge transformation of photon $i$ takes the form $\varepsilon_{i\mu} \rightarrow \varepsilon_{i\mu} + \lambda k_{i\mu}$ for an arbitrary constant $\lambda$. Under this the vertex changes as
\begin{align}
	V_{\eta}^{x'x}[k_i,\varepsilon _i] \rightarrow  V_{\eta}^{x'x}[k_i,\varepsilon _i] +i\lambda \int_0 ^T d \tau _i \, \partial _{\tau _i} \text{e}^{ik_i \cdot x(\tau _i)}= V_{\eta}^{x'x}[k_i,\varepsilon _i] +i\lambda \left(  \text{e}^{ik_i \cdot x'}- \text{e}^{ik_i \cdot x}   \right)\, .
\label{eqDeltaV}
\end{align}
Here the last term -- that depends only upon the endpoints of the trajectory -- does not contribute to the on-shell matrix elements by the Ward identity (once Fourier transformed to momentum space, the exponential factors shift the location of the poles away from the mass shell so they cannot contribute in the LSZ formula). Moreover, for an external photon attached to a closed fermion loop, the total derivative integrates to zero. This means that the non-trivial gauge transformation properties of the propagator come only from the transformation of the internal photons\footnote{The plane wave decomposition of the external photons also enters in the prefactor of (\ref{eqSGordon}), yet it has been shown in \cite{fppaper2} that these terms do not contribute on-shell, for the same reason that they lack the correct LSZ poles; as such their gauge transformation need not be considered here.} which we go on to determine in the following section.

\subsection{Variation of the propagator}
\label{secProp}
Firstly we re-derive the original LKF transformation of the two-point function using the worldline techniques presented above. For this analysis, we split (\ref{eqSxi}) into two parts
 \begin{equation}
	S(x; x' | \xi) =  [m+i\Dspg]\big\langle K^{x'x}[A^{\gamma} + \bar{A}]\big\rangle_{\bar{A},\, \xi} - \big\langle e\bar{\As}(x') K^{x'x}[A^{\gamma} + \bar{A}]\big\rangle_{\bar{A},\, \xi}.  
\label{eqprop}
\end{equation}
After the gauge transformation the first term will produce the multiplicative LKF law seen in the scalar case, (\ref{eqScalarLKFTFinal}), plus an additional, unwanted derivative of this exponential factor. This extra derivative term will be cancelled by the non-multiplicative part of the gauge transformation of the second term. 

It will be convenient to define the more general $\xi$-dependent functional 
\begin{align}
\mathcal{I}[J,M;\xi) := \Braket{\text{e}^{ie \int d^Dx \,J[x] \cdot \bar{A}[x]} \prod _{j=1}^M K_j ^{x'_{\pi(j)}x_j} [A^{\gamma} + \bar{A}]}_{\bar{A},\,  \xi},
\label{eqIxi}
\end{align}
which can be used to generate insertions of $\bar{A}$ through functional differentiation. To evaluate this we require the path integral representation of the kernel, (\ref{eqKPathInt}). It is clear that the functional integral over $\bar{A}$ will then be Gaussian. However, further simplifications can be engendered by taking advantage of a supersymmetry in the worldline action. The action is invariant under the transformations
\begin{equation}
	\delta x^{\mu} = -2\zeta\psi^{\mu}\, , \qquad \delta \psi^{\mu} = \zeta \dot{x}^{\mu}\,,
\end{equation} 
with $\zeta$ a constant Grassmann number, which motivates us to formulate the worldline theory in superspace. So we extend our parameter domain to $1$-$1$ superspace, ${\tau \rightarrow \tau\, |\, \theta}$, by introducing the Grassmann parameter $\theta$. We can then define the superfield and super-derivative
\begin{align}
	\mathbb{X}^{\mu}(\tau,\theta) &= x^{\mu}(\tau) + \sqrt{2} \theta (\psi ^{\mu}(\tau) + \eta ^{\mu}) \\
	\mathbb{D} &= \partial _{\theta} - \theta \partial _{\tau}.
	\end{align}
Integrals of superfields over the whole of superspace, such as $\int d\tau \int d\theta \, \mathbb{X}$, are invariant (up to boundary terms) under supersymmetric transformations. In particular, we can express (\ref{eqKPathInt}) as
\begin{equation}
K^{x'x}[A] = 	2^{-\frac{D}{2}} \symb \int_{0}^{\infty}dT \, \e^{-m^{2}T} \int \mathscr{D}\mathbb{X}\,\e^{-S_{0}[\mathbb{X}]- S_{\textrm A}[\mathbb{X}]}
\label{eqKSuper}
\end{equation}
in which appear the free particle action, $S_{0}[\mathbb{X}]$, and the interaction with the gauge field, $S_{\textrm{A}}[\mathbb{X}]$, which (up to total derivatives) can be written in superspace as
\begin{align}
	S_{0}[\mathbb{X}] &= \int_{0}^{T}d\tau \int d\theta \Big[-\frac{1}{4}\mathbb{X}\cdot \mathbb{D}^{3}\mathbb{X}\Big]\\
	S_{\textrm A}[\mathbb{X}] &= \int_{0}^{T}d\tau \int d\theta \Big[-ie A[\mathbb{X}]\cdot \mathbb{D}\mathbb{X}\Big]\, .
\end{align}
The boundary conditions on $\mathbb{X}$ are inherited from $x$ and $\psi$.

To determine $\mathcal{I}$ we decompose the gauge field into external and internal photons. The path integral over the potential of the internal photons, $\bar{A}$, is then determined by using the superspace representation of the kernels in $\mathcal{I}$ and completing the square to arrive at
\begin{align}
\mathcal{I}[J,M;\xi) =& \prod _{j=1}^M 2^{-\frac{D}{2}} \text{symb}^{-1} \int _0^{\infty}\!dT_j\, \e^{-m^2  T_j}\int \!\mathscr{D} \mathbb{X}_{j}  \,\e^{-\sum _{l=1}^M S^{(l)}_{0\gamma}[\mathbb{X}_{l}] - S_{i}[\mathbb{X}, J]}\, ,
\label{eqIintegrated}
\end{align}
where $S_{0,\gamma}^{(l)}[\mathbb{X}]$ consists of the free action for trajectory $l$ along with its coupling to the external photons and we have defined the generalised interaction term
\begin{align}
	S_{i}[\mathbb{X}, J] &= \frac{e^2}{2} \iint d^D y d^D y' \, \mathscr{J}(y) \cdot G(y-y';\xi) \cdot \mathscr{J}(y')\, ,\\
\mathscr{J}^{\mu}(y) &= J^{\mu}(y) + \sum _{l=1}^M \int _0^{T_l} d \tau _l \int d\theta _l \, \delta ^D (y - \mathbb{X}_l) \mathbb{D}_l \mathbb{X}_l^{\mu}. 
\label{eqJ}
\end{align}
Note that in the current case, the inverse symbol map must first order the variables $\eta_{l}$ in ascending order to reproduce the numeration of variables in the product in (\ref{eqIxi}) \textit{before} converting them into products of $\gamma$-matrices.

The crucial observation is that, using (\ref{eqGConfig}), a change in the gauge parameter ${\xi \rightarrow \xi + \Delta \xi}$ causes a variation in $S_{i}$ that is again a total derivative:
\begin{align}
	\hspace{-1em}\mathscr{J}(y) \cdot \Delta _{\xi} G \cdot \mathscr{J}(y') = \frac{\Delta \xi}{16 \pi ^{\frac{D}{2}}} \Gamma \left[  \frac{D}{2} - 2   \right] \mathscr{J}(y) \cdot \partial_y \mathscr{J}(y') \cdot \partial _{y'} [(y-y')^2]^{2- \frac{D}{2}}\, .
\label{eqgaugechange}
\end{align}
Using this in the variation of (\ref{eqIintegrated}) we get the gauge variation of the exponent divided into three term. The contribution independent of the external source, $J$, coincides with (\ref{eqDeltaS}) from the scalar case according to
\begin{align}
		\Delta _{\xi} S^{(k,l)}_{i\pi} = -&\Delta \xi \frac{e^2}{32 \pi ^{\frac{D}{2}}} \Gamma \left( \frac{D}{2} -2   \right) \int_0^{T_k} d \tau _k \int_0^{T_l} d \tau _l \int d \theta _k \int d \theta _l\,  \mathbb{D}_k \mathbb{D}_l [  ( \mathbb{X}_k - \mathbb{X}_l ) ^2  ]^{2- \frac{D}{2}} \nonumber \\
		=& \Delta \xi \frac{e^2}{32 \pi ^{\frac{D}{2}}} \Gamma \left( \frac{D}{2} -2   \right) \int_0^{T_k} d \tau _k \int_0^{T_l} d \tau _l \, \partial _{\tau _k} \partial _{\tau _l} [  ( x_k - x_l ) ^2  ]^{2- \frac{D}{2}}\, .
\label{eqLKFTSpin}
\end{align} 
This corresponds to the transformation of the propagator caused by a change of gauge in the internal photon propagators that couple to the worldline trajectories. Aside from this there are two terms in (\ref{eqgaugechange}) involving the source that provide the term denoted by $\Delta _{\xi}I_M$ in \cite{UsLKFT1}. It can be split up into the sum of  
\begin{align}
\Delta_{\xi}I_{M}^{(1)}&= - e^2 \sum _{i=1}^M \int d^D y \int _0 ^{T_i} d \tau _i \int d \theta _i \ J[y] \cdot G(y-\mathbb{X}_i; \xi) \cdot \mathbb{D}_i \mathbb{X}_i \nonumber  \\
	&= \frac{\Delta \xi e^2}{16 \pi ^{\frac{D}{2}}} \Gamma \left[ \frac{D}{2} - 2 \right] \sum _{i=1}^M \int _0 ^{T_i} d \tau _i \int d^D x\,  J(x) \cdot \partial _x \partial _{\tau _i} [(x-x_i)^2]^{2- \frac{D}{2}} \, ,
\end{align}
and
\begin{align}
\Delta _{\xi}I_{M}^{(2)}&= - \frac{e^2}{2} \int d^D y \int d^D z \ J[y] \cdot G(y-z;\xi) \cdot J[z] \nonumber \\
 &= - \frac{\Delta \xi e^2}{32 \pi ^{\frac{D}{2}}} \Gamma \left[ \frac{D}{2} - 2 \right] \iint d^D x d^D x'\,  J(x) \cdot \partial _{x} J(x') \cdot \partial _{x'} [(x-x')^2]^{2-\frac{D}{2}}\, .
\end{align}
Put together, these imply that $\mathcal{I}$ transforms with the simple multiplicative law
\begin{align}
\mathcal{I}[J,M;\xi + \Delta \xi) = \mathcal{I}[J,M;\xi) \text{e}^{- \sum_{k,l=1}^M \Delta _{\xi} S_{i\pi}^{(k,l)} + \Delta _{\xi} I_M}\, ,
\end{align}
which generalises slightly the original LFK transformation.

With this, we can analyse how the propagator (\ref{eqprop}) transforms. We can express it in terms of $\mathcal{I}$ as
\begin{equation}
S(x;x'|\xi) =  [m + i \slashed{D}^{' \gamma}] \mathcal{I}[0,1; \xi) + i \frac{\delta}{\delta \slashed{J}'} \mathcal{I}[1,1;\xi) \Big| _{J=0}.
\end{equation}
Making a transformation of the gauge parameter, we have
\begin{align}
S(x;x'|\xi + \Delta \xi) =  [m + i \slashed{D}^{' \gamma}]\big( \mathcal{I}[0,1; \xi)  \text{e}^{- \Delta _{\xi} S_i}\big) + i \frac{\delta}{\delta \slashed{J}'} \big(\mathcal{I}[1,1;\xi)  \text{e}^{- \Delta _{\xi} S_i + \Delta _{\xi} I_1} ]\big)\Big| _{J=0},
\label{eqSDeltaXi}
\end{align}
\noindent where both partial and functional derivatives act through onto the exponential factors. However,   the two terms which arise from applying the derivatives to the exponents cancel due to the general relation
\begin{align}
  \frac{\delta}{\delta \slashed{J}(x_i')} \Delta _{\xi} I^{(1)}_M  = \slashed{\partial}^{\prime}_{i} \sum _{k,l=1}^M \Delta _{\xi} S^{(k,l)}_{i \pi}\, ,
\end{align}
and the fact that $\frac{\delta}{\delta \slashed{J}(x_i')} \Delta _{\xi} I^{(2)}_M = 0$ when $J = 0$. This allows for the exponential factor in the first term of (\ref{eqSDeltaXi}) to be commuted to the left, resulting in the following transformation for the propagator 
\begin{align}
S(x;x'|\xi + \Delta \xi) = \text{e}^{- \Delta _{\xi} S_{i}} S(x;x'| \xi)\, ,
\end{align}
which corresponds to the original LKF transformation. It takes the same form as in the scalar case, as  obtained in the derivation of Landau and Khalatnikov that is independent of the details of the matter field. 

\section{$N$-point functions}
\label{secNPoint}
The main contribution of this paper is to provide additional details that prove the generalisation of the LKF transformation of the propagator to arbitrary correlators, expanding upon the results reported in \cite{UsLKFT1}. To this end we generalise the propagator, that corresponds to the field theory correlator $\langle \bar{\Psi}(x)\Psi(x^{\prime})\rangle$, to the correlator of an arbitrary even number, $N = 2n$, of fields, $\langle \bar{\Psi}(x_{1})\cdots \bar{\Psi}(x_{n}) \Psi(x^{\prime}_{1})\cdots \Psi(x^{\prime}_{n})\rangle$.  This $N$-point function can be decomposed into partial amplitudes	
	\begin{align}
	S(x_1 \ldots x_n ; x'_{1} \ldots x'_{n} | \xi ) = \sum_{\pi \in S_n} \mathcal{S}_{\pi}(x_1 \ldots x_n ;x'_{\pi (1)} \ldots x'_{\pi (n)} | \xi), \label{SNpoint}
	\end{align}	  
where the partial amplitude $\mathcal{S}_{\pi}$ represents the contribution in which the field $\Psi(x_{\pi(i)}^{\prime})$ is contracted with the conjugate field $\bar{\Psi}(x_{i})$, defined as 
\begin{align}
	\mathcal{S}_{\pi}(x_1 \ldots x_n; x'_{\pi (1)} \ldots x'_{\pi (n)} | \xi ) =  \braket{[m + i \slashed{D}'_1]K_1^{x'_{\pi (1)} x_1} \cdots [m + i \slashed{D}'_n] K_n^{x'_{\pi (n)} x_n }}_{\bar{A} ,\,  \xi}\, . 
\label{eqSMultiple}
\end{align}
The generalised LKF transformation will be determined in terms of the transformation of these partial amplitudes. We shall express the kernels in terms of path integrals over particle trajectories; as discussed above, the gauge transformation due to the external photons is fixed by the Ward identity, so we are again free to focus on the variation induced by changing the gauge parameter of the internal, virtual photons. 

In this section we continue to apply function methods to evaluate correlators using path integrals; a complementary approach that focuses more on the combinatorial aspects is developed in appendix \ref{AppExp}. We begin with an intermediate result that extends $\mathcal{I}$ of (\ref{eqIxi}). We consider the gauge transformation of the following function (for $M \geqslant 1)$:
	\begin{align}
		\Braket{[m + i\slashed{D}'_1] \text{e}^{ie \int d^D x \, J[x] \cdot \bar{A}[x]} \prod _{i=1}^M K^{x_i' x_i}}_{\bar{A}, \xi}\, . 
		\label{eqJ1}
\end{align}	 
Applying the change $\xi \rightarrow \xi + \Delta \xi$, our analysis follows the previous case except that we no longer set $J = 0$ at the end of the calculation. It is straightforward to verify that
	\begin{align}
	&\Braket{[m + i\slashed{D}'_1]\text{e}^{ie \int d^D x \, J[x] \cdot \bar{A}[x]} \prod _{i=1}^M K^{x_i' x_i}}_{\bar{A}, \xi + \Delta \xi} \nonumber \\
	&= \text{e}^{- \sum _{k,l}^M  \Delta _{\xi} S_i^{(k,l)} +  \Delta _{\xi} I_M  }\Braket{\big[m + i\slashed{D}'_1 + i \slashed{\partial}'(\Delta _{\xi} I^{(2)}_{M})\big] \text{e}^{ie \int d^D x \,J[x] \cdot \bar{A}[x]} \prod _{i=1}^M K^{x_i' x_i}}_{\bar{A}, \xi} \, .
	\end{align}
This will be used below to prove the transformation rule of the $N$-point correlator.

to this end it is useful to consider a slightly more general functional, defined as (again ${\pi \in S_{M}}$)
\begin{align}
	\mathcal{J}[J,K,M;\xi) := \Braket{\text{e}^{ie\int d^Dx\,  J[x] \cdot \bar{A}[x]} \prod _{i=1}^K [m + i\slashed{D}'_i] \prod _{j=1}^M K_j^{x'_{\pi(j)}x_j}[A^{\gamma} + \bar{A}]}_{\bar{A},\xi}, \ \ \ K \leq M\, . \label{eqJcal}
\end{align}
When we use the path integral representation of the various kernels the symbol map continues to order the Grassmann variables according to the two products in $\mathcal{J}$, before converting them back to $\gamma$-matrices. In fact, we should stress that the ordering of the $\gamma$-matrices in $\mathcal{J}[J,K,M;\xi)$ does not yet correspond to that in the partial $N$-point amplitude but we shall see that this is easily remedied under the symbol map.

We claim that $\mathcal{J}$ transforms in the following way:
\begin{align}
	&\mathcal{J}[J,K,M;\xi + \Delta \xi) = \Big[ \mathcal{J}[J,K,M;\xi) + \sum_{k=0}^{K-1}\prod _{l=1}^k [m + i \widehat{\slashed{D}}'_l](i \slashed{\partial}' _{k+1} \Delta_{\xi}I_M^{(2)})\mathcal{J}^{(k+2)}[J,K,M;\xi) \Big] \text{e}^{-\sum _{k,l=1}^M \Delta _{\xi} S_{i\pi}^{(k,l)} + \Delta_{\xi}I_M},
	\label{eqJtransformation}
\end{align}
with $\widehat{D}'_{\mu} = \partial _{\mu}' + \frac{\delta}{\delta J^{\mu}(x')} + ieA_{\mu}(x')^{\gamma}$ a generalised differential operator, whose derivatives act through onto everything to their right. The superscript in $\mathcal{J}^{(k+2)}[J,K,M;\xi)$ indicates that the variable $i$ in (\ref{eqJcal}) runs from $k+2$ to $K$. The proof is most easily done by induction on $K$ with the results for (\ref{eqIxi}) and (\ref{eqJ1}) validating the base cases corresponding to $K = 0$ and $K = 1$ respectively. Given this transformation, we examine the case $\mathcal{J}[J,K+1,M;\xi+ \Delta \xi)$ (maintaining $K+1 \leq M$). We generate the additional insertion of $\bar{A}$ that appears in this case by functional differentiation with respect to $J$, which leads to:
\begin{align}
  \mathcal{J}[J,K+1,M;\xi+ \Delta \xi) &= [m + i\widehat{\slashed{D}}'_1] \mathcal{J}^{(2)}[J,K+1,M;\xi+ \Delta \xi) \nonumber \\ 
  &= [m + i\widehat{\slashed{D}}'_1] \Big[ \mathcal{J}^{(2)}[J,K+1,M;\xi)  \nonumber \\
   &+ \sum_{k=0}^{K}\prod _{l=1}^k \big[m + i \widehat{\slashed{D}}'_l\big](i \slashed{\partial}' _{k+1} \Delta_{\xi}I_M^{(2)})\mathcal{J}^{(k+2)}[J,K+1,M;\xi) \Big] \e^{-\sum _{k,l=1}^M \Delta _{\xi} S_{i\pi}^{(k,l)} + \Delta_{\xi}I_M}.
    \label{eqJDerivDeltaXi}
\end{align}
Distributing the various derivatives in $\widehat{\slashed{D}}'_1$, the first term on the right hand side gives
\begin{align}
	 [m + i\widehat{\slashed{D}}'_1]  \mathcal{J}^{(2)}[J,K+1,M;\xi) &=  \mathcal{J}[J,K+1,M;\xi) \text{e}^{-\sum _{k,l=1}^M \Delta _{\xi} S_{i\pi}^{(k,l)} + \Delta_{\xi}I_M} \nonumber \\
	 &+ \mathcal{J}^{(2)}[J,K+1,M;\xi)\big(i \dsp_{1} + \frac{\delta}{\delta \Js^{\prime}}\big) \e^{-\sum _{k,l=1}^M \Delta _{\xi} S_{i\pi}^{(k,l)} + \Delta_{\xi}I_M}\, .
\end{align} 
In the second line there are various cancellations between the derivatives: the term proportional to $\slashed{\partial}'\Delta _{\xi} S_{i\pi}^{(k,l)}$ cancels with the term involving $\frac{\delta}{\delta \slashed{J}'}\Delta _{\xi}I_M^{(1)}$ (as seen in the previous section), and the term with $\slashed{\partial}' \Delta _{\xi}I_M^{(1)}$ cancels against the piece proportional to $\frac{\delta}{\delta \slashed{J}'} \Delta _{\xi} I^{(2)}_M$ due to an analogous relation
\begin{align}
\frac{\delta}{\delta \slashed{J}(x'_i)} \Delta _{\xi} I_M^{(2)} = \slashed{\partial}'_i \Delta _{\xi} I_M^{(1)}.
\end{align}
The surviving term from the second line, $(i\dsp_{1}\Delta_{\xi}I_{M}^{(2)}) \mathcal{J}^{(2)}[J,K+1,M;\xi)\e^{-\sum _{k,l=1}^M \Delta _{\xi} S_{i\pi}^{(k,l)} + \Delta_{\xi}I_M}$, combines with the variation produced by the second line of (\ref{eqJDerivDeltaXi}) and allows us to write 
\begin{align}
\mathcal{J}[J,K+1,M;\xi+ \Delta \xi)&= \mathcal{J}[J,K+1,M;\xi) \e^{-\sum _{k,l=1}^M \Delta _{\xi} S_{i\pi}^{(k,l)} + \Delta_{\xi}I_M} \nonumber \\ 
&+ (i\slashed{\partial}'_1\Delta _{\xi}I^{(2)}_M) \mathcal{J}^{(2)}[J,K+1,M;\xi) \text{e}^{-\sum _{k,l=1}^M \Delta _{\xi} S_{i\pi}^{(k,l)} + \Delta_{\xi}I_M} \nonumber \\
  &+  [m + i\widehat{\slashed{D}}'_1]\left[ \sum_{k=1}^{K}\prod _{l=2}^k [m + i \widehat{\slashed{D}}'_l](i \slashed{\partial}' _{k+1} \Delta_{\xi}I_M^{(2)})\mathcal{J}^{(k+2)}[J,K+1,M;\xi) \right] \text{e}^{-\sum _{k,l=1}^M \Delta _{\xi} S_{i\pi}^{(k,l)} + \Delta_{\xi}I_M} \nonumber \\
 &= \text{e}^{-\sum _{k,l=1}^M \Delta _{\xi} S_{i\pi}^{(k,l)} + \Delta_{\xi}I_M} \mathcal{J}[J,K+1,M;\xi) \nonumber \\
 &+ \Big[ \sum_{k=0}^{K}\prod _{l=1}^k [m + i \widehat{\slashed{D}}'_l](i \slashed{\partial}' _{k+1}\Delta_{\xi}I_M^{(2)})\mathcal{J}^{(k+2)}[J,K+1,M;\xi) \Big] \text{e}^{-\sum _{k,l=1}^M \Delta _{\xi} S_{i\pi}^{(k,l)} + \Delta_{\xi}I_M}.
\end{align}
where the $k = 0$ contribution to the sum comes from the second line of the first expression and we have made a relabelling to begin the product at $l = 1$ (we also recall that the derivatives $\widehat{\slashed{D}}_{l}$ act on everything to their right). This shows that $\mathcal{J}[J,K,M;\xi)$ transforms as claimed in (\ref{eqJtransformation}) for all $K$. 

\subsection{Generalised LKF transformation}
We can use this immediately to derive the generalised LKF transformation rule for the $(N=2n)$-point partial amplitude. To do this, we first observe that the $\gamma$-matrices play no role in determining the functional form of the transformation: we can either factorise them outside of the expectation value or ask that the symbol map reorders the Grassmann variables, $\eta_{i}$, under the path integral accordingly -- we only have to ensure that we return to the initial ordering of these matrices by the end of the calculation. Choosing the ordering that corresponds to the propagator, (\ref{eqSMultiple}), we fix $K=M=n$ and evaluate the transformation just derived on $J = 0$ which gives
\begin{align}
\hspace{-1em}&\mathcal{S}_{\pi}(x_1 \ldots x_n; x'_{\pi (1)} \ldots x'_{\pi (n)} | \xi + \Delta \xi ) =  \text{e}^{-\sum _{k,l=1}^n \Delta _{\xi} S_{i\pi}^{(k,l)}} \mathcal{S}_{\pi}(x_1 \ldots x_n; x'_{\pi (1)} \ldots x'_{\pi (n)} | \xi )  \nonumber \\
\hspace{-1em}& \ \ \ +  \Big[ \sum_{k=0}^{n-1}\prod _{l=1}^k [m + i \widehat{\slashed{D}}'_l](i \slashed{\partial}' _{k+1} \Delta_{\xi}I_n^{(2)})\mathcal{J}^{(k+2)}[J,n,n;\xi) \Big] \text{e}^{-\sum _{k,l=1}^n \Delta _{\xi} S_{i\pi}^{(k,l)} + \Delta_{\xi}I_n} \Big| _{J=0}.
\end{align}
Now we assert that $J = 0$ actually kills the second term on the right hand side of this result. This can be seen by noting that $\Delta_{\xi}I_{n}^{(2)}$ is quadratic in $J$, so that we need to apply two function derivatives to it to obtain something that survives this limit. The resulting expression will depend only on the two variables of these derivatives, neither of which will be the same as the partial derivative which also acts on $\Delta_{\xi} I^{(2)}_n$. Thus the vanishing of this second term when we take $J=0$ results in equation (\ref{eqSLKFT}),
\begin{align}
&\mathcal{S}_{\pi}(x_1 \ldots x_n; x'_{\pi (1)} \ldots x'_{\pi (n)} | \xi + \Delta \xi )= T_{n} \mathcal{S}_{\pi}(x_1 \ldots x_n; x'_{\pi (1)} \ldots x'_{\pi (n)} | \xi )\, ,
\end{align} 
where we have followed the notation of \cite{LKFTWL1, LKFTWL2} in defining
\begin{equation}
	T_{n}:= \text{e}^{-\sum _{k,l=1}^n \Delta _{\xi} S_{i\pi}^{(k,l)}}\,.
\end{equation}
Thus we have arrived at the generalized LKF transformation for the $N$-point correlation function for spinor QED, which has turned out to be the same transformation as in the scalar case.

We can add to the discussion in \cite{LKFTWL1, LKFTWL2} with the observation that, after summing over $k$ and $l$, the complete LKF factor, $T_{n}$, is in fact \textit{independent} of the permutation, $\pi$, that fixes the partial amplitude. Instead, $T_{n}$ is a function only of the endpoints of the $(N = 2n)$ worldlines, since the sum forces all of these endpoints to pair up in all possible combinations. As an important consequence, not only do the partial amplitudes determine the LKF transformation, but they all transform in the same way, leading immediately to the multiplicative result for the propagator itself, with a \textit{global} transformation
\begin{equation}
	S(x_1 \ldots x_n ; x'_{1} \ldots x'_{n} | \xi + \Delta \xi) = T_{n} S(x_1 \ldots x_n ; x'_{1} \ldots x'_{n} | \xi )\, , 
\end{equation}
where $T_{n}$ can be determined using, say, the $\Delta_{\xi}S_{iI}^{(k, l)}$ corresponding to the identity permutation -- as claimed in equation (\ref{eqSLKFTGlobal}). Note that as a consequence an analogous statement holds for scalar QED (with the same prefactor that is now promoted to a global multiplicative factor). We refer to appendix \ref{AppExp} for an alternative method for deriving the result proved here.
\subsubsection{Specific examples: Conformal cross ratios}
Now that we have arrived at the generalised LKF transformation, it is worthwhile to consider some examples. Since the results coincide with those of scalar QED, this analysis also expands upon the examples given in \cite{LKFTWL1, LKFTWL2}. In general, the complete, or non-perturbative Green functions may have poles, potentially to all orders, in the physical dimension, $D = D_{0}$, in which cases we use dimensional regularisation, fixing $D = D_{0} - 2\epsilon$. The LKF factor, $T_{n}$, should then be taken to all orders in $\epsilon$, which can be a non-trivial task. Here we restrict our attention to the first non-trivial contributions in various dimensions.

We begin with the case $D_{0} = 4$, also discussed in \cite{LKFTWL1, LKFTWL2}. The $\frac{1}{\epsilon}$ pole in the gamma functions cancels between the terms of $\Delta S_{i \pi}^{(k, l)}$ so that $T_{n}$ is finite in the limit $\epsilon \to 0$. In this case we have
\begin{equation}
	T_{n} = \left(\prod_{k, l = 1}^{n} r_{\pi}^{(k, l)}\right)^{\frac{\Delta \xi e^{2}}{32 \pi^{2}}} + \mathcal{O}(\epsilon)\, ,
\end{equation}
with, as in \cite{LKFTWL1, LKFTWL2}, $r_{\pi}^{(k, l)}$ the conformal cross ratio corresponding to the endpoints of the lines with labels $k$ and $l$:
\begin{equation}
	r_{\pi}^{(k, l)} \equiv \frac{(x_{k} - x_{l})^{2}(x^{\prime}_{\pi(k)} - x^{\prime}_{\pi(l)})^{2}}{(x_{k} - x^{\prime}_{\pi(l)})^{2}(x^{\prime}_{\pi(k)} - x_{l})^{2}}\, .
\end{equation}
Note that this factor appears regardless of the mass of the spinor particle propagating between these endpoints (which can occur because this factor does not depend upon the details of the propagation between these points). The original LKF transformation is recovered by setting $ n = 1 $, for which 
\begin{equation}
	T_{1} = \bigg[\frac{(x - x)^{2}(x^{\prime} - x^{\prime})^{2}}{((x - x^{\prime})^{2})^{2}}\bigg]^{\frac{\Delta \xi \alpha}{8 \pi}} + \mathcal{O}(\epsilon)\,.
\end{equation}
Finally we replace the vanishing numerator by the cut-off $((x_{0})^{2})^{2}$ to arrive at (\ref{eqSLKF4}). Repeating this trick for the arbitrary correlator and including the contributions from all endpoints we arrive at the simplification
\begin{equation}
	\prod_{k, l = 1}^{n}r_{\pi}^{(k, l)} = (x_{0}^{2})^{2n}\, \frac{\prod_{l > k = 1}^{n}\big((x_{k} - x_{l})^{2}(x^{\prime}_{k} - x^{\prime}_{l})^{2}\big)^{2}} {\prod_{k, l = 1}^{n} ((x_{k} - x^{\prime}_{l})^{2})^{2}}\, ,
\end{equation}
which gives the leading order contribution to the LKF factor independently of the permutation defining the partial amplitude. It corresponds to the product of (regulated) conformal cross ratios of the endpoints of $n$ lines for all possible pairings of initial and final points.

In $D_{0} = 3$ dimensions the LKF factor is without poles and we get an order unity contribution
\begin{equation}
	\Delta_{\xi}S_{i\pi}^{(k, l)} = - \frac{\Delta \xi e^{2}}{16 \pi} \Big[ \big| x_{k} - x_{l} \big| - \big| x_{k} - x^{\prime}_{\pi(l)} \big| - \big|x^{\prime}_{\pi(k)} - x_{l}\big| + \big| x^{\prime}_{\pi(k)} - x^{\prime}_{\pi(l)} \big| \Big]\,.
\end{equation}
Summing over $k$ and $l$ the LKF exponent can be simplified to	
\begin{equation}
	-\sum_{k, l = 1}^{n}\Delta_{\xi}S_{i\pi}^{(k, l)} = \frac{\Delta \xi e^{2}}{8 \pi} \Big\{ \sum_{l > k = 1}^{n} \Big[ \big| x_{k} - x_{l} \big| + \big| x^{\prime}_{k} - x^\prime_{l} \big| \Big] - \sum_{k, l = 1}^{n} \big| x_{k} - x^{\prime}_{l} \big| \Big\}\, ,
\end{equation}
which is now manifestly independent of the permutation $\pi$. Again, fixing $n = 1$, we find
\begin{equation}
	T_{1} = \e^{-\frac{\Delta \xi \alpha}{2} | x - x' |} + \mathcal{O}(\epsilon)\, ,
\end{equation}
which has given (\ref{eqSLKF3}). We repeat that the divergences in loop diagrams would require evaluation of higher order terms in $\epsilon$ we do not consider here.

For the case $D_{0} = 2$, $\Delta_{\xi}S_{i\pi}^{(k, l)}$ again has a pole in $\epsilon$. The singular part is (we introduce an arbitrary mass scale $\mu$ for dimensional consistency)
\begin{align}
	\Delta_{\xi}S_{i\pi}^{(k, l)}\big|_{\epsilon^{-1}} &= \frac{\Delta \xi e^{2} \mu^{2}}{32 \pi \epsilon} \Big[(x_{k} - x_{l})^{2} - (x_{k} - x^{\prime}_{\pi(l)})^{2} - (x^{\prime}_{\pi(k)} - x_{l})^{2} + (x^{\prime}_{\pi(k)} - x^{\prime}_{\pi(l)})^{2} \Big]\nonumber \\
	&=-\frac{\Delta \xi e^{2} \mu^{2}}{16 \pi \epsilon} (x_{k}-x^{\prime}_{\pi(k)})\cdot (x_{l}-x^{\prime}_{\pi(l)})\, .
\end{align}
Summing this over values of $l$ and $k$ removes the dependence on the permutation, giving
\begin{equation}
	-\sum_{k, l =1}^{n}\Delta_{\xi}S_{i\pi}^{(k, l)}\big|_{\epsilon^{-1}} = \frac{\Delta \xi e^{2} \mu^{2}}{16 \pi \epsilon} \Big[\sum_{k=1}^{n} (x_{k} - x^{\prime}_{k}) \Big]^{2}\,.
\end{equation}
For the original LKF transformation, with $n = 1$, the result is trivial:
\begin{equation}
	T_{1}\big|_{\epsilon^{-1}} = \e^{ \frac{\Delta \xi \alpha \mu^{2}}{4 \epsilon} (x - x^{\prime})^{2}}\,.
	\label{eqT12m1}
\end{equation}
We also give the finite contribution for this case. There are two contributions to $\Delta_{\xi}S$,
\begin{align}
	\Delta_{\xi}S_{i\pi}^{(k, l)}\big|_{\epsilon^{0}} &= \frac{\Delta \xi e^{2} \mu^{2}}{32 \pi}\bigg\{ (1- \gamma_{E} ) (x_{k}-x^{\prime}_{\pi(k)})\cdot (x_{l}-x^{\prime}_{\pi(l)})\nonumber \\
	&+ \Big[(x_{k} - x_{l})^{2}\log\big[\pi \mu^{2} (x_{k} - x_{l})^{2} \big] - (x_{k} - x^{\prime}_{\pi(l)})^{2}\log\big[\pi \mu^{2} (x_{k} - x^{\prime}_{\pi(l)})^{2} \big] \nonumber \\
	&- (x^{\prime}_{\pi(k)} - x_{l})^{2}\log\big[\pi \mu^{2} (x^{\prime}_{\pi(k)} - x_{l})^{2} \big] + (x^{\prime}_{\pi(k)} - x^{\prime}_{\pi(l)})^{2}\log\big[\pi \mu^{2} (x^{\prime}_{\pi(k)} - x^{\prime}_{\pi(l)})^{2} \big] \Big] \bigg\}\, ,
\end{align}
where $\gamma_{E}$ is the Euler–Mascheroni constant. Summing over $k$ and $l$ we find
\begin{align}
	-\sum_{k, l =1}^{n}\Delta_{\xi}S_{i\pi}^{(k, l)}\big|_{\epsilon^{0}} &= \frac{\Delta \xi e^{2} \mu^{2}}{16 \pi  } \bigg\{ (\gamma_{E} - 1) \Big[\sum_{k=1}^{n} (x_{k} - x^{\prime}_{k}) \Big]^{2} + \sum_{k, l = 1}^{n} (x_{k} - x^{\prime}_{l})^{2} \log\big[\pi \mu^{2}(x_{k} - x^{\prime}_{l})^{2} \big] \nonumber \\
	&- \sum_{l > k = 1}^{n} \left[  (x_{k} - x_{l})^{2} \log\big[\pi \mu^{2}(x_{k} - x_{l})^{2} \big] + (x^{\prime}_{k} - x^{\prime}_{l})^{2} \log\big[\pi \mu^{2}(x^{\prime}_{k} - x^{\prime}_{l})^{2} \big]\right] \bigg\}\,,
\end{align}
in which we have removed the dependence on the permutation $\pi$. The simplest case of $n=1$ gives the original transformation at constant order,
\begin{equation}
	T_{1}\big|_{\epsilon^{0}} = \e^{ \frac{\Delta \xi \alpha \mu^{2}}{4 } (x - x^{\prime})^{2} \big[ \log[\pi \mu^{2}(x - x^{\prime})^{2}] + \gamma_{E} - 1 \big] }\,.
	\label{eqT120}
\end{equation}
Combining this with (\ref{eqT12m1}) it is important to note that this time the pole is \textit{not} cancelled when $\Delta_{2-2\epsilon}(0)$ is subtracted, which implies an essential singularity for the LKF transformation, (\ref{eqLKFTProp}), in the limit $\epsilon \rightarrow 0$. In a perturbative calculation, the poles, of arbitrary order, would thus need to be taken into account in the transformation. This is consistent with the observation of \cite{AShok} that in $D=2$ the pole of the fermion propagator is not gauge invariant in covariant gauges at any finite order in perturbation theory.
These considerations, along with the physically interesting aspects of two dimensional QED, make further studies of this case of both theoretical and practical interest for ongoing and future work.

\section{Perturbation theory}
\label{secPert}
It is clear by now that the LKF transformation is non-perturbative in nature. However, in a practical perturbative calculation one would like to see how it works order by order in the loop expansion. For this purpose, in this section we consider a specific fixed loop-order process as an example to illustrate the gauge transformation of the internal photon propagators and represent the transformation diagrammatically. Since the generalised LKF transformation has turned out to be the same as in the scalar case, we can rework the perturbative discussion in \cite{LKFTWL1, LKFTWL2} for the present case. In particular, although it is possible to obtain gauge dependent higher-loop order contributions from a given amplitude, we restrict attention here to the transformation of terms at a fixed loop order.

\begin{figure}[h]
  \centering
    \includegraphics[width=0.5\textwidth]{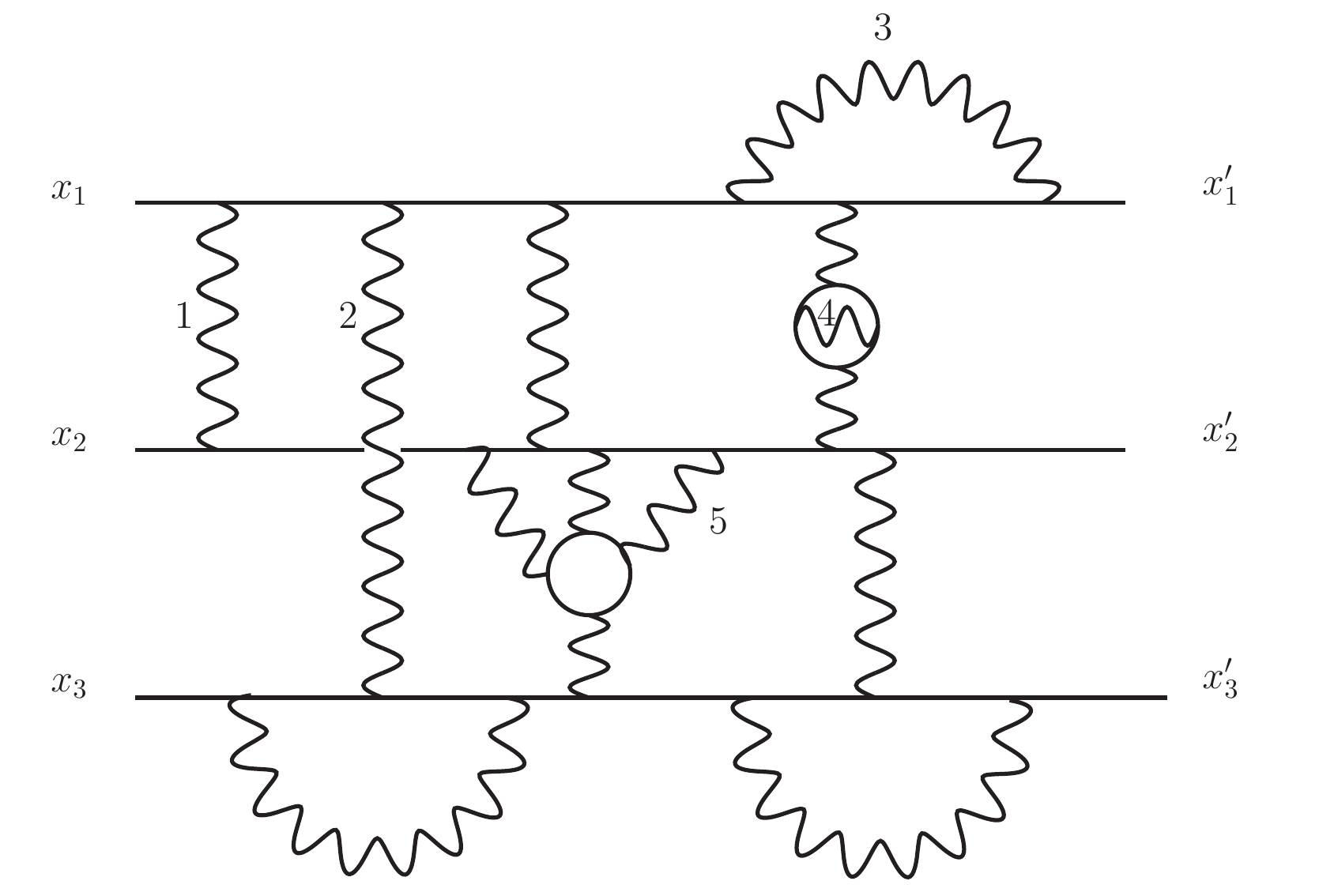}
     \caption{A Feynman diagram for a typical process with three fermion propagators representing six external particle legs at twelve loop order in configuration space. Here $x_i$ and $x'_i$ are the end-points of the propagators ($i=1,2,3$). The numbers $1-5$ indicate some of the photons that could be gauge transformed.}
     \label{figLKFT}
\end{figure}

For instance consider the Feynman diagram depicted in Fig. {\ref{figLKFT}	} with three electron propagators and twelve loops. It should be understood that we consider the sum of this diagram together with all the ones that differ from it only by {\it letting photon legs slide along spinor lines}. This is a very complicated process but here we are interested in the application of the LKF transformation to some of the internal photons which are indicated with numbers. We recall that the gauge transformation properties of an amplitude are determined completely by the photons exchanged between two fermion lines, or along one fermion line (like photons $1,2,3$); in Fig. \ref{figLKFT} photon $4$ and $5$ do not produce a gauge transformation because they start and / or end on an electron loop. 

An advantage of our formalism is that we have the freedom to effect a change of gauge parameter on \textit{individual photons} in isolation, which affects the amplitude according to (\ref{eqLKFTSpin}): it converts the photon connecting two propagators with endpoints $x_{l}$ and $x_{k}$ into a multiplicative factor of $-\Delta_\xi S_{i\pi}^{(k,l)}$. Thus the gauge transformation of a photon eliminates that photon and leaves a diagram of lower loop order (the appropriate factor of the coupling constant is contained in $\Delta_{\xi}S_{\pi}^{(k, l)}$). 
 
\begin{figure}
    \centering
    \begin{subfigure}[b]{0.315\textwidth}
        \includegraphics[width=\textwidth]{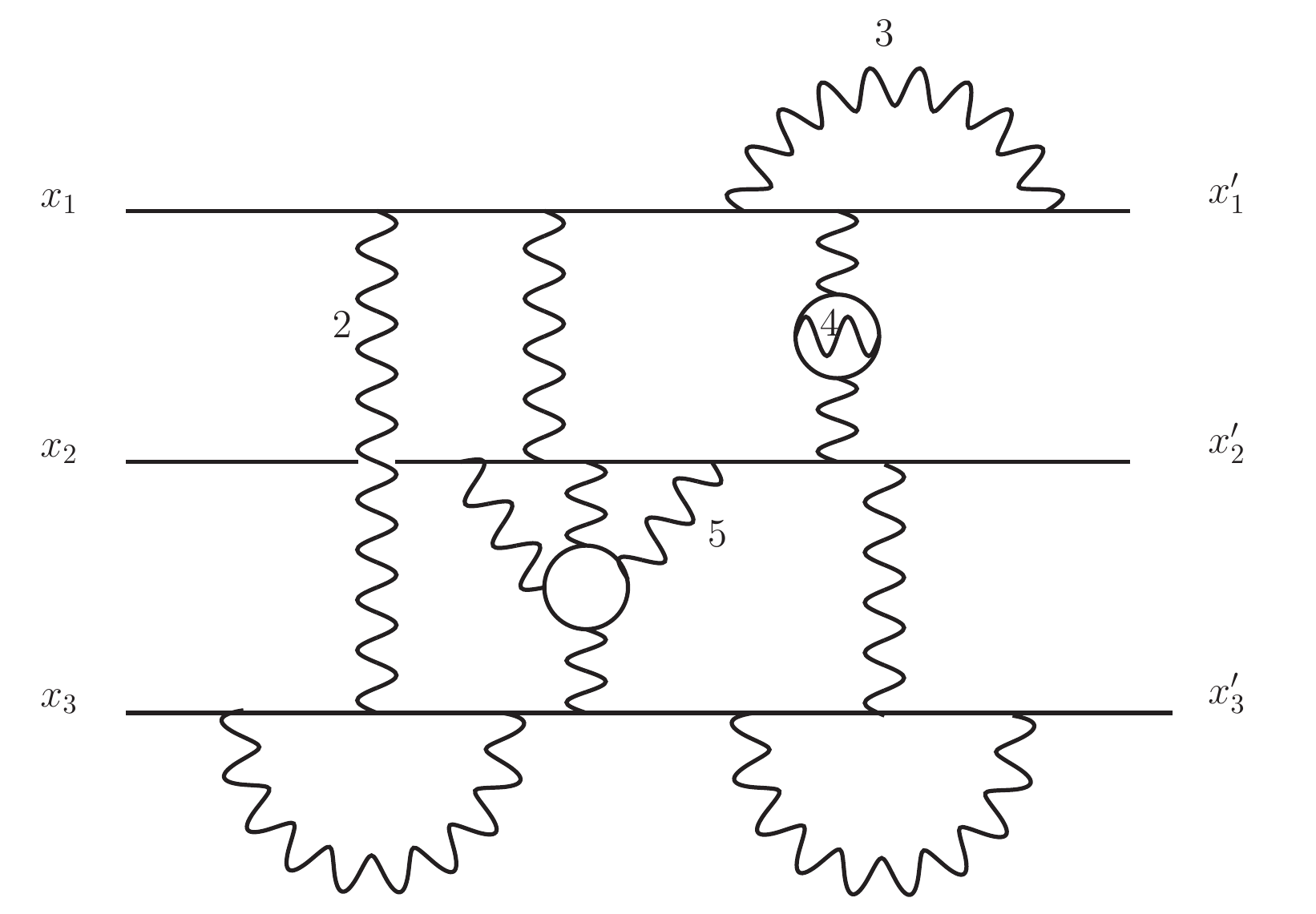}
        \caption{\tiny Gauge transformation of photon $1$.}
        \label{lkft-spin1}
    \end{subfigure}
    ~ 
    \begin{subfigure}[b]{0.315\textwidth}
        \includegraphics[width=\textwidth]{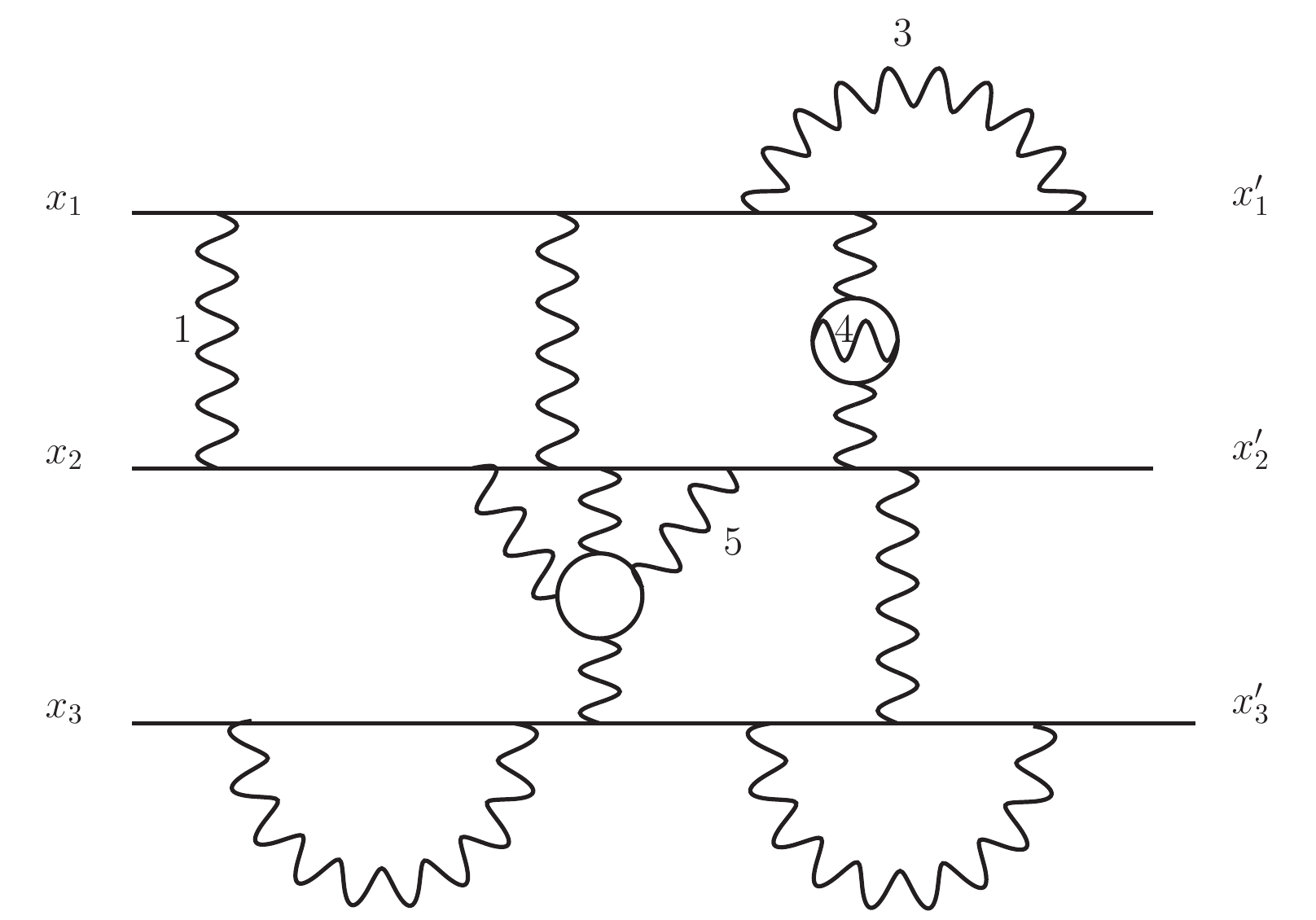}
        \caption{\tiny Gauge transformation of photon $2$.}
        \label{lkft-spin2}
    \end{subfigure}
    ~ 
    \begin{subfigure}[b]{0.315\textwidth}
        \includegraphics[width=\textwidth]{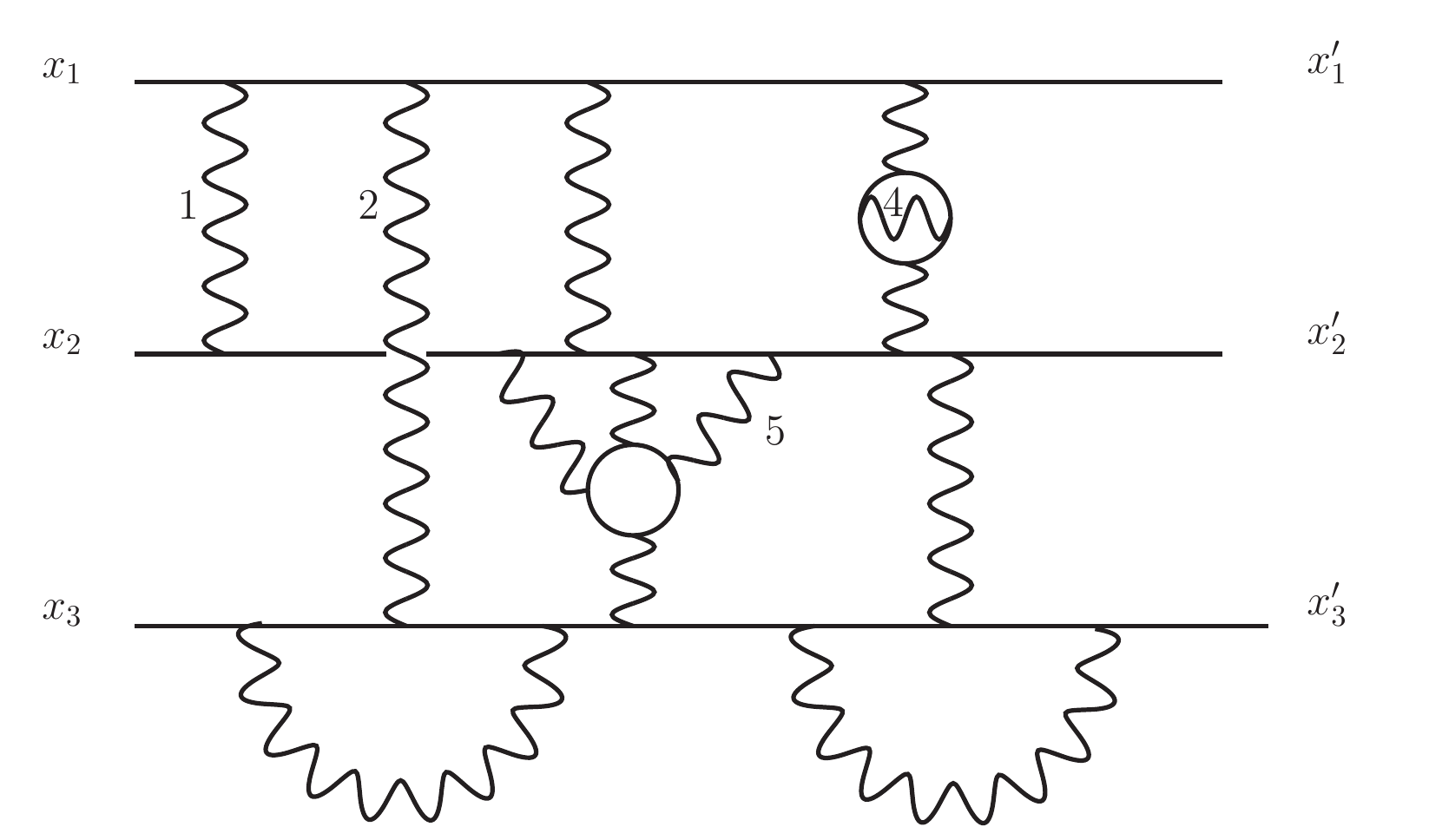}
        \caption{\tiny Gauge transformation of photon $3$.}
        \label{lkft-spin3}
    \end{subfigure}
      \begin{subfigure}[b]{0.35\textwidth}
        \includegraphics[width=\textwidth]{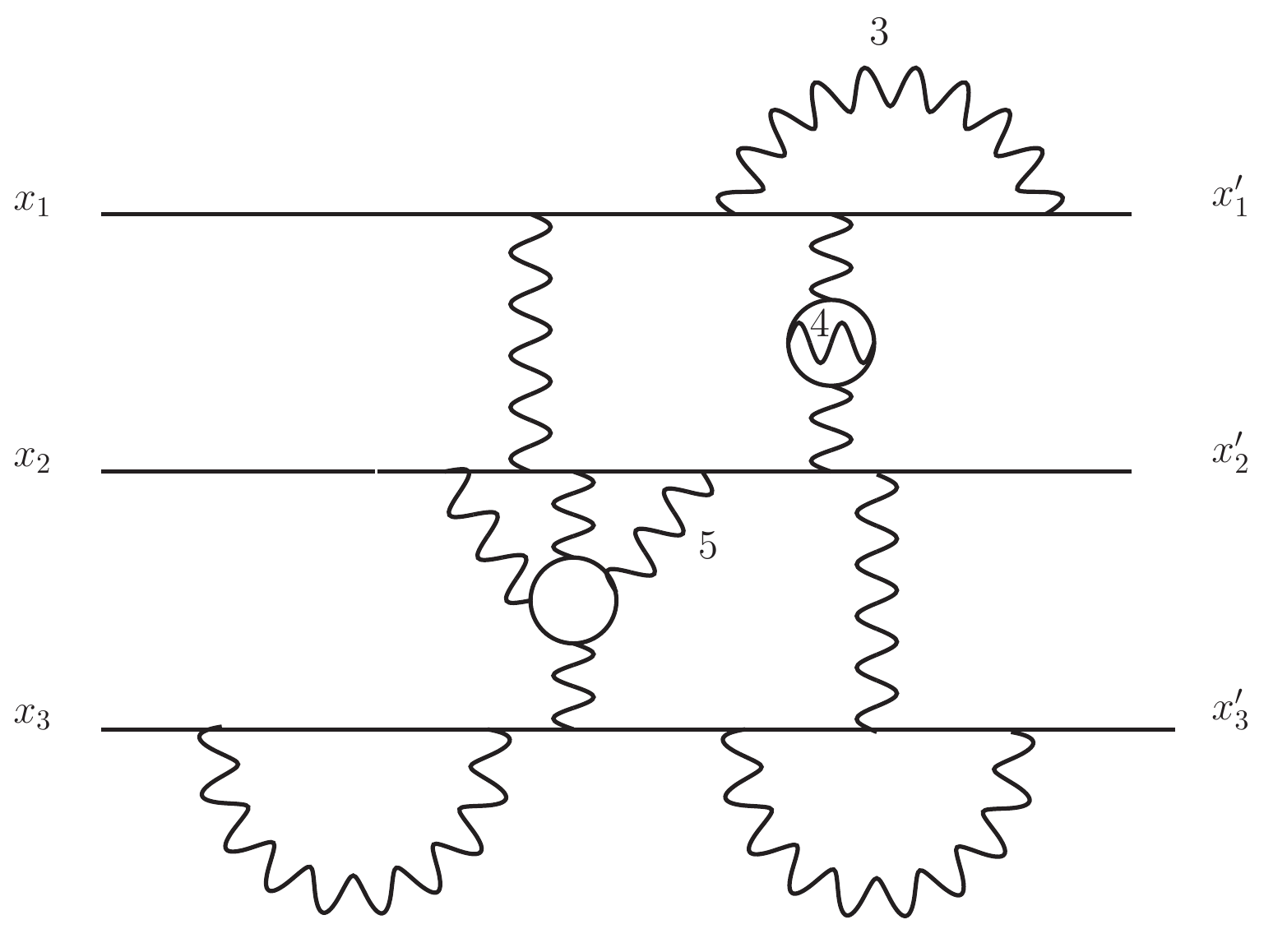}
        \caption{\tiny Simultaneous gauge transformation of photon $1$ and $2$.}
        \label{lkft-spin12}
    \end{subfigure}
     \begin{subfigure}[b]{0.35\textwidth}
        \includegraphics[width=\textwidth]{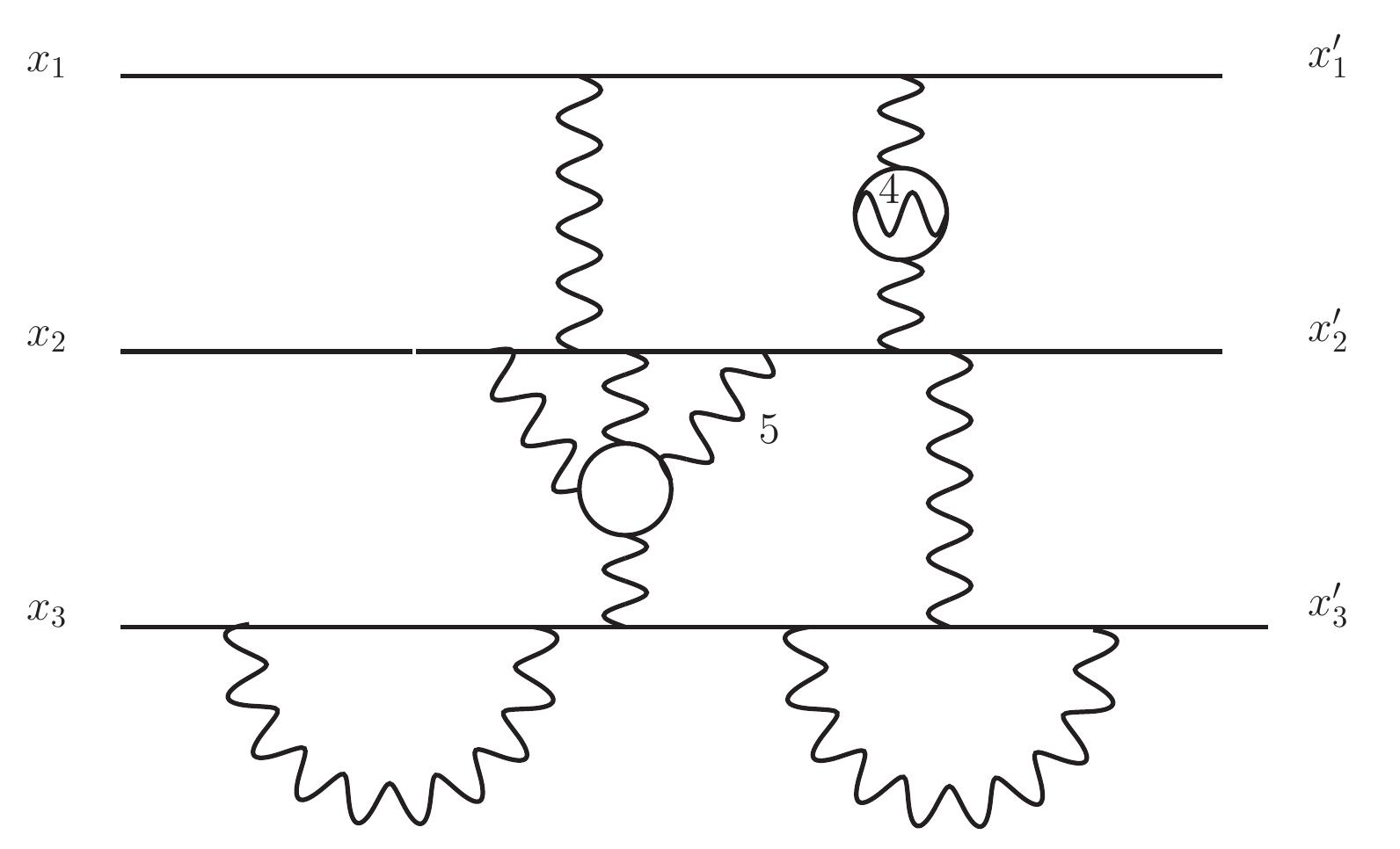}
        \caption{\tiny Simultaneous gauge transformation of photon $1$, $2$ and $3$.}
        \label{lkft-spin123}
    \end{subfigure}
    $+\cdots $
    \caption{Diagrammatic presentation of gauge transformation of internal photons one by one or of some of them simultaneously using the generalised LKF transformations in perturbation theory.}
    \label{figDiagrams}
\end{figure}

If we apply the LKF to the internal photons in Fig. \ref{figLKFT} we obtain the gauge variation of this diagram (see also Fig. \ref{figDiagrams}): 
\bear
\Delta_\xi\, {\rm Fig} \,\ref{figLKFT}&=&(-2\Delta_\xi S_{i\pi}^{(1,2)}) {\rm Fig}\,\ref{lkft-spin1}+(-2\Delta_\xi S_{i\pi}^{(1,3)}) {\rm Fig}\,\ref{lkft-spin2}+(-\Delta_\xi S_{i\pi}^{(1,1)}) {\rm Fig}\,\ref{lkft-spin3}+\cdots\non
&&+(-2\Delta_\xi S_{i\pi}^{(1,2)}) \,(-2\Delta_\xi S_{i\pi}^{(1,3)}) {\rm Fig}\,\ref{lkft-spin12}+\cdots\non
&&+(-2\Delta_\xi S_{i\pi}^{(1,2)}) \,(-2\Delta_\xi S_{i\pi}^{(1,3)})\,(-\Delta_\xi S_{i\pi}^{(1,1)}) {\rm Fig}\,\ref{lkft-spin123}+\cdots\non
&&+\,\vdots
\ear
In the above equation the first line represents the gauge transformation of individual photons, the second line is for the simultaneous transformation of pairs of photons, the last line for the simultaneous gauge transformation of three photons and so on which is extremely straightforward using the above LKF rules. 

Although this section has focussed on perturbation theory in configuration space, it is also possible to transfer the LKF transformations found here to the perturbative expansion in momentum space. This has been achieved for the propagator in scalar and spinor QED in $D = 3$ and $D = 4$ dimensions \cite{AdnanRaya, ScalarPerturb, LKFPert}; the generalisation we have developed here will allow us to apply these techniques to arbitrary correlation functions in future work. In particular, in a further publication we shall deal with the non-trivial pole structure in the two dimensional LKF factor in the context of momentum space perturbation theory.  
 
\section{Conclusion}
\label{secConc}
We have applied first quantised techniques to determine the transformation of arbitrary fermion correlation functions induced by varying the linear covariant gauge parameter of virtual photons that give loop corrections to the free correlators. These coordinate space transformations generalise the original studies of Landau, Khalatnikov and Fradkin for the propagator to the general case of the $N$-point functions of spinor QED and are completely non-perturbative.	 We recover the original result as a special case with $N = 2$. 

The generalised transformations were found by studying the variation induced in partial amplitudes that pair up initial and final points in a particular way. Their variation turned out to be the same as that of their counterparts in scalar QED and corresponds to the introduction of total derivatives in worldline parameter integrals. However, we noted here that the functional form of this variation does not depend upon the ordering implied by the partial amplitude and as such factorises out of the sum over orderings. We were thus led to a simple, multiplicative transformation for the complete correlators in both scalar and spinor QED which is the natural generalisation of the multiplicative transformation for the propagator. 

It was important to check this, since the multiplicative form of the transformation could have been broken,  even at the level of the partial amplitudes, by the derivative structure of the worldline representation of the correlators, an issue raised in \cite{LKFTWL1}. We have manifested that there is a precise cancellation between all such terms that means that the multiplicative form is maintained after all. In the main text this was shown using functional methods; a more direct proof is given in the appendix.  We strongly suspect that finding this factorisation would be substantially more difficult using standard techniques. 

Since the transformation takes the same form as in scalar QED, its application in perturbation theory is the same as in the former case, originally worked out in \cite{LKFTWL1, LKFTWL2} and discussed briefly in section \ref{secPert} above. In the case of four dimensional QED the first order expansion of the multiplicative factor is written in terms of conformal cross ratios of the correlation function arguments. We have added to that work by considering the transformation in two-dimensional QED wherein there is a divergence that affects the LKF factor to all orders in the perturbative expansion. As such one can expect poles of arbitrary order to enter the correlation functions after being non-perturbatively gauge transformed.

Current and future work in this context will develop the perturbative application further, in particular to analyse the momentum space transformation of the propagator for the two-dimensional case as a massive analogue \cite{QED21} of the Schwinger model \cite{SchwingerMod1, SchwingerMod2}. Here there will be the additional difficulty of the pole to treat carefully. Given the extension of the LKF transformation considered here, we shall also be able to examine the momentum space transformation of higher order Green functions in four dimensional QED. Likewise, in lower dimensions the transformations of the propagator in \cite{RQED1, RQED3} could now be extended to higher order correlation functions.

Whilst we have worked entirely within quantum electrodynamics, the worldline techniques we have applied can be adapted to non-Abelian theories to study the gauge transformation implied by virtual gluons. Moreover, the worldline formalism extends to a gravitational background which would allow for studies of the diffeomorphism structure of the propagator or correlation functions due to virtual graviton exchange. Likewise, the gauge structure of more complicated objects such as the propagator in an electromagnetic background, or interaction vertices can be studied with an aim to obtain information about their form factor decomposition. Such work would have application in informing analyses of the Schwinger-Dyson equations by supplying further restrictions on solutions that incorporate the gauge structure implied by the LKF transformations uncovered here.

\subsection*{Acknowledgements}
The authors thank Adnan Bashir for sharing his insight on LKFTs and their application to the Schwinger-Dyson equations and for recommending various references. They are grateful to Pietro Dall'Olio for helpful discussions. JN and JPE are supported by CONACyT. JPE acknowledges funding from CIC-UMICH.
\appendix
\section*{Appendices}
\section{Expectation values}
\label{AppExp}
Here we define how correlation functions of the quantum background field, $\bar{A}$, are calculated and provide an alternative proof of the cancellation of undesired derivative terms against the contributions from the $\bar{\As}$ terms in the prefactors of (\ref{eqSGordon}). Throughout this appendix we shall set the value of the charge to $e = 1$.

We define the expectation value $\langle \prod_{i = 1}^{N} \bar{A}_{\mu_{i}}(x_{i}) \rangle$ according to the Euclidean space path integral over $\bar{\textrm{A}}$ with a particular gauge fixing action $S_{\textrm{gf}}(\xi) = -\int d^{4}x (\partial \cdot \bar{A})^{2} / (2\xi)$ that imposes the covariant linear gauge with parameter $\xi$:
\begin{equation}
	\big\langle \prod_{i = 1}^{N} \bar{A}_{\mu_{i}}(x_{i}) \big\rangle_{\xi} := \int \mathscr{D}\bar{A}(x) \,  \prod_{i = 1}^{N} \bar{A}_{\mu_{i}}(x_{i}) \e^{-\int d^{D}x \left[-\frac{1}{4} \bar{F}_{\mu\nu}\bar{F}^{\mu\nu}\right] - S_{\textrm{gf}}(\xi)}\,.
\end{equation}
In particular the two-point function reproduces the Green function (\ref{eqGConfig}), $\langle  \bar{A}_{\mu}(x) \bar{A}_{\nu}(x') \rangle_{\xi} = G_{\mu\nu}(x - x'; \xi)$. 

As usual the insertions of the prefactors $\bar{A}_{\mu_{i}}(x_{i})$ can be generated by functional differentiation with respect to a source term, $S[J] = i\int d^{D}x\, J(x) \cdot \bar{A}(x)$. We can take advantage of this to give an alternative derivation of the result given in the main text regarding the cancellation of the unwanted derivatives of $\Delta_{\xi}S_{i\pi}$ that would otherwise spoil the multiplicative form of the LKF transformations. To this end we consider
\begin{equation}
	\mathcal{K}_{\pi}(n, N; \xi) := \big\langle \prod_{i=1}^{n}\bar{\As}_{i}(x_{i})\prod_{j=1}^{N}K^{x^{\prime}_{\pi(j)}x_{j}}_{j}[\bar{A}]\big\rangle_{\xi} ~; \qquad \pi \in S_{N}\, ,
\end{equation}
where we have temporarily ignored the external photons which play no role in this calculation. The subindices on the $\bar{\As}$ are reminders that the $\gamma$ matrices must be placed in the correct order according to the product in (\ref{eqSMultiple}) but it will become clear below that this is not important for now. Expressing the $K_{j}$ in their path integral representation, (\ref{eqKSuper}), and generating the $\bar{\As}$ by functional differentiation we arrive at
\begin{align}
	\hspace{-2em} \mathcal{K}_{\pi}(n, N; \xi) = \symb \Big\{& \prod_{j = 1}^{N}2^{-\frac{D}{2}}\int_{0}^{\infty} dT_{j} \, \text{e}^{-m^2T_j}\int\mathscr{D}\mathbb{X}(\tau_{j}, \theta_{j})\nonumber \\
	 &	\e^{-\sum_{l = 1}^{N}S_{0}^{l}[\mathbb{X}_{l}]} \frac{\delta^{n}}{\delta \Js^{n}} \big\langle  \e^{i \sum_{i = 1}^{N} \int d\tau_{i} \int d\theta_{i} \mathbb{D}_{i} \mathbb{X}_{i} \cdot \bar{A}(\mathbb{X}_{i}) + i\int d^{D}x J(x) \cdot \bar{A}(x) } \big\rangle_{\xi} \Big\}\, ,
	\label{eqIPathInt}
\end{align}
where we abbreviate $\frac{\delta^{n}}{\delta \Js^{n}} := \frac{(-i)^{n}\delta^{n}}{\delta \Js(x_{1}) \cdots \delta \Js(x_{n})}\big|_{J = 0}$. Completing the square in the final exponent allows for the expectation value to be computed which supplies
\begin{equation}
	 \mathcal{K}_{\pi}(n, N; \xi) = \symb \Big\{  \prod_{j = 1}^{N}2^{-\frac{D}{2}}\int_{0}^{\infty}  dT_{j}\, \text{e}^{-m^2T_j}\int\mathscr{D}\mathbb{X}(\tau_{j}, \theta_{j}) \e^{-\sum_{l = 1}^{N}S_{0}^{l}[\mathbb{X}_{i}]} \frac{\delta^{n}}{\delta \Js^{n}}\e^{\frac{1}{2} \int d^{D}y d^{D}y' \mathcal{J}(y) \cdot G(y - y'; \xi) \cdot \mathcal{J}(y')}\Big\}
	\label{eqIGxi}
\end{equation}
where the current is $\mathcal{J}^{\mu}(y) = J^{\mu}(y) + \sum_{i = 1}^{N}\int d\tau_{i}\int \, d\theta_{i} \delta^{D}(y - \mathbb{X}_{i}) \mathbb{D}_{i}\mathbb{X}^{\mu}_{i}$. We are not interested in the precise form of this result, nor the path integrals over the $\mathbb{X}_{i}$ but rather how this quantity changes when we vary $\xi$. As such we realise the change of gauge directly in the Green function so that the integrand of (\ref{eqIGxi}) changes according to
\begin{align}
	\hspace{-3.5em}\frac{\delta^{n}}{\delta \Js^{n}} \big\langle  \e^{i\sum_{i = 1}^{N} \int d\tau_{i} \int d\theta_{i} \mathbb{D}_{i} \mathbb{X}_{i} \cdot \bar{A}(\mathbb{X}_{i}) + i\int d^{D}x J(x) \cdot \bar{A}(x) } \big\rangle_{\xi + \Delta \xi} &= \sum_{m = 0}^{n} \sum_{\textrm{perm}\, \{\rho_{m}\}}\frac{\delta^{m}_{\rho}}{\delta \Js^{m}} \e^{\frac{1}{2}\int d^{D}x \int d^{D}x' \mathcal{J}(x) \cdot \Delta_{\xi} G(x - x') \cdot \mathcal{J}(x')} \nonumber \\
	& \times \frac{\delta^{n-m}_{\rho}}{\delta \Js^{n-m}} \big\langle  \e^{i \sum_{i = 1}^{N} \int d\tau_{i} \int d\theta_{i} \mathbb{D}_{i} \mathbb{X}_{i} \cdot \bar{A}(\mathbb{X}_{i}) + i\int d^{D}x J(x) \cdot \bar{A}(x) } \big\rangle_{\xi},
	\label{eqDeltaAA}
\end{align}
where the sum over partitions, $\{\rho_{m}\}$, counts all ways to choose $m$ variables from $\{1, \ldots n\}$ which appear in the first functional derivative -- the remaining $n-m$ are then placed in the second. Here the exponent of the first term on the right hand side is derived from the $\xi$-dependent parts of (\ref{eqGConfig}) and can be decomposed as shown in the main text:
	\begin{align}
\hspace{-2em}	\frac{1}{2}\int d^{D}x \int d^{D}x' &\mathcal{J}(x) \cdot \Delta_{\xi} G(x - x') \cdot \mathcal{J}(x') = -\sum_{i, j = 1}^{N} \Delta_{\xi}S_{i\pi}^{(i, j)} + \Delta_{\xi}I_{N}^{(1)} + \Delta_{\xi}I_{N}^{(2)}
	\label{eqDeltaAAExp}
\end{align}
where we have used (\ref{eqgaugechange}) and computed the integrals over the $\theta_{i}$ as in section \ref{secProp}.

The first term in (\ref{eqDeltaAAExp}) generates the global exponential factor common to all terms in the Green functions. Meanwhile the $m$ functional derivatives in (\ref{eqDeltaAA}) will act on the functions $J(x)$ in the $\Delta_{\xi}I_{N}$ of (\ref{eqDeltaAAExp}) to produce various terms. Since the derivatives are contracted into $\gamma$-matrices the structure is particularly simple. We must choose $k$ pairs and $l$ singletons such that $2k + l = m$. Each pair produces an insertion of the form
\begin{equation}
	\frac{\Delta \xi }{16 \pi^{\frac{D}{2}}}\Gamma\Big[\frac{D}{2} - 2\Big] \gamma \cdot \partial_{x_{i}} \gamma \cdot \partial_{x_{j}}\left[\left(x_{i} - x_{j})\right)^{2}\right]^{2-\frac{D}{2}}
\end{equation}
which we recognise as $\slashed{\partial}_{x_{i}}\slashed{\partial}_{x_{j}} \Delta_{\xi}S_{i\pi}^{(i, j)}$. Similarly, the singletons produce factors
\begin{equation}
	\frac{\Delta \xi }{32 \pi^{\frac{D}{2}}}\Gamma\Big[\frac{D}{2} - 2\Big]\sum_{i=1}^{N}\int_{0}^{T_{i}}d\tau_{i}\int d^Dx \, \gamma \cdot \partial_{x_{j}} \frac{\partial}{\partial \tau_{i}}\left[\left(x_{j} - x(\tau_{i})\right)^{2}\right]^{2-\frac{D}{2}}
\end{equation}
which is of course $\sum_{i = 1}^{N}\slashed{\partial}_{x_{j}}\Delta_{\xi} S_{i\pi}^{(i, j)}$. As such (\ref{eqDeltaAA}) can be written as
\begin{align}
\hspace{-2.5em}	\frac{\delta^{n}}{\delta \Js^{n}} \big\langle  &\e^{i \sum_{i = 1}^{N} \int d\tau_{i} \int d\theta_{i} \mathbb{D}_{i} \mathbb{X}_{i} \cdot \bar{A}(\mathbb{X}_{i}) + i\int d^{D}x J(x) \cdot \bar{A}(x) }\big\rangle_{\xi + \Delta \xi} = \nonumber \\
\hspace{-2.5em}		&\hspace{0em}\sum_{m = 0}^{n}\sum_{\textrm{perm}\, \{\rho_{m}\}} \frac{\delta^{n-m}_{\rho}}{\delta \Js^{n-m}} \big\langle  \e^{i \sum_{i = 1}^{N} \int d\tau_{i} \int d\theta_{i} \mathbb{D}_{i} \mathbb{X}_{i} \cdot \bar{A}(\mathbb{X}_{i}) + i\int d^{D}x J(x) \cdot \bar{A}(x) } \big\rangle_{\xi} \nonumber \\
	\hspace{-3em}	&\hspace{0.5em} \times \sum_{2k + l = m} \sum_{\sigma \in S_{m}}\,\prod_{i=1}^{k} \left( \slashed{\partial}_{x_{\sigma(2i-1)}}\slashed{\partial}_{x_{\sigma(2i)}} \Delta_{\xi}S_{i\pi}^{(\sigma(2i-1), \sigma(2i))} \right)\prod_{j = 2k+1}^{m}\left(\sum_{p = 1}^{N}\slashed{\partial}_{x_{\sigma(j)}}\Delta_{\xi} S_{i\pi}^{(p, \sigma(j))}\right)\e^{-\sum_{r, s = 1}^{N}\Delta_\xi S_{i\pi}^{(r, s)}},
	\label{eqDeltaAAS}
\end{align}
where the $\{\sigma\}$ permute the set $\{\rho(1), \ldots \rho(m)\}$. Finally we note that differentiating $\sum_{r, s = 1}^{N}\Delta_\xi S_{i\pi}^{(r, s)}$ twice with respect to distinct positions, $x_{1}$ and $x_{2}$ leaves a function only of $x_{1}$ and $x_{2}$ so that a further derivative with respect to any $x_{3}$ kills the result. For this reason we recognise that the final line of (\ref{eqDeltaAAS}) is precisely the action of $m$ derivatives $\slashed{\partial}_{x}$ on the global exponent.
Substituting this into (\ref{eqIPathInt}) we arrive at
\begin{align}
	\hspace{-1em}\mathcal{K}_{\pi}(n, N; \xi + \Delta \xi) &= \sum_{m = 0}^{n} \sum_{\textrm{perm}\, \{\rho_{m}\}} \Big\langle \prod_{j=m+1}^{n} \bar{\As}(x_{\rho(j)}) \prod_{k = 1}^{N}K_{k}^{x^{\prime}_{\pi(k)}x_{k}}	 \Big\rangle_{\xi}  \left[\prod_{i=1}^{m}i \slashed{\partial}_{x_{\rho(i)}} \right]   \e^{-\sum_{l,m=1}^{N} \Delta_{\xi}S_{i\pi}^{(l, m)}} \,.
\label{eqDeltaXiAs}
\end{align}

In fact we can improve this to give an iterative formula relating changes in the various matrix elements. We define a new ``difference'' operator, $\diff_{\xi}$, which returns the non-multiplicative terms in the transformations of matrix elements as follows:
\begin{equation}
	\dif_{\xi}\langle \cdots \rangle_{\xi} = \langle \cdots \rangle_{\xi + \Delta \xi} - \langle \cdots \rangle_{\xi}\e^{-\Delta_{\xi}S}
\label{eqDif}
\end{equation}
where we have denoted $\Delta_{\xi}S = \sum_{l, m}\Delta_{\xi}S_{i \pi}^{(l, m)}$, allowing the case that the sum is empty. To give some examples,
\begin{align}
	\dif_{\xi}\Big\langle  \prod_{j = 1}^{N}K_{j}^{x^{\prime}_{\pi(j)}x_{j}}	 \Big\rangle_{\xi} &= 0\\
	\dif_{\xi}\Big\langle  \bar{\As}_{1}\prod_{j = 1}^{N}K_{j}^{x^{\prime}_{\pi(j)}x_{j}}	 \Big\rangle_{\xi} &= \Big\langle   \prod_{j = 1}^{N}K_{j}^{x^{\prime}_{\pi(j)}x_{j}}	 \Big\rangle_{\xi} i\ds_{1}\e^{-\sum_{l,m=1}^{N} \Delta_{\xi}S_{i\pi}^{(l, m)}} \\
	&= \dif_{\xi}\Big \langle i\ds_{1}  \prod_{j = 1}^{N}K_{j}^{x^{\prime}_{\pi(j)}x_{j}}	 \Big\rangle_{\xi}
	\label{eqDifEx}
\end{align}
From these relations follows immediately the LKF transformation for the propagator, which in this notation takes the form
\begin{equation}
	\dif_{\xi}\Big \langle \big[ m + i\ds^{\prime} - \bar{\As}(x^{\prime}_{1})\big] K_{1}^{x^{\prime}, x} \Big\rangle = 0.
\end{equation}
To include derivatives we note that there is a commutator $[\diff_{\xi}, \partial_{\mu}] = \partial_{\mu}\e^{-\Delta_{\xi}S}$ which follows by direct computation from the definition (\ref{eqDif}). This allows a nice way of arriving at the last line of (\ref{eqDifEx}).  

With this notation and using the properties of $\diff_{\xi}$ we can convert (\ref{eqDeltaXiAs}) into  a stronger statement that is equivalent to the LKFT. This requires moving the derivatives acting on the LKF exponent inside of the expectation value. To achieve this, we first prove a general property: with $\mathscr{A}(\bar{A})$ any function of $\bar{A}$ we have, for $n \geqslant 1$ 
\begin{align}
	\hspace{-2em }\big \langle \mathscr{A} \prod_{k=1}^{N}K_{k}^{x'_{\pi(k)} x_{k}} \big \rangle &i\dsp_{1} i \dsp_{2} \ldots i \dsp_{n} \e^{-\Delta_{\xi}S} = (-1)^{n-1}\Big\{ \dif_{\xi} \big \langle i\dsp_{1} i \dsp_{2} \ldots i \dsp_{n} \,  \mathscr{A} \prod_{k=1}^{N}K_{k}^{x'_{\pi(k)} x_{k}} \big \rangle \nonumber \\ 
	\hspace{-2em }&-\sum_{m=1}^{n} \sum_{\{\rho_{m}\}} (-1)^{m-1} i\dsp_{\rho(1)} \ldots i \dsp_{\rho(m)} \dif_{\xi} \big \langle i\dsp_{\rho(m+1)} \ldots i\dsp_{\rho(n)}\,   \mathscr{A} \prod_{k=1}^{N}K_{k}^{x'_{\pi(k)} x_{k}} \big \rangle \Big\}\, ,
	\label{eqDerivs}
\end{align} 
where the sum over $\{\rho_{m}\}$ is over the selection of $m$ out of the $n$ derivatives. Although this very general result follows from repeated application of the commutator, it is quicker to use induction. For $n = 1$ we may repeat the same steps in (\ref{eqDifEx}) with the additional insertion of $\mathscr{A}$, since the commutator holds; the $n = 2$ case, which is needed to check the alternating sign, is discussed below. Assuming, then, that the relation holds for $n$ derivatives we write $\big \langle \mathscr{A} \prod_{k=1}^{N}K_{k}^{x'_{\pi(k)} x_{k}} \big \rangle i\dsp_{1} i \dsp_{2} \ldots i \dsp_{n+1} \e^{-\Delta_{\xi}S}$ as
\begin{align}
	\hspace{-2em }i \dsp_{n+1} \Big[ \big \langle \mathscr{A} \prod_{k=1}^{N}K_{k}^{x'_{\pi(k)} x_{k}} \big \rangle i\dsp_{1} i \dsp_{2} \ldots i \dsp_{n} \e^{-\Delta_{\xi}S} \Big]- \big \langle i \dsp_{n+1} \mathscr{A}\prod_{k=1}^{N}K_{k}^{x'_{\pi(k)} x_{k}} \big \rangle i\dsp_{1} i \dsp_{2} \ldots i \dsp_{n} \e^{-\Delta_{\xi}S}
\end{align}
where in the second term the derivative $\dsp_{n+1}$ does not act beyond the expectation value. The inductive hypothesis immediately shows that these two terms simply split up contributions in which $\dsp_{n+1}$ is outside or inside of the expectation value respectively and hence give the result for $n+1$ derivatives. 

We can use this immediately in (\ref{eqDeltaXiAs}) to move derivatives inside of the expectation values. Then the LKFT corresponds to the fact that all of the terms coming from the second line of (\ref{eqDerivs}), involving derivatives of the variations, cancel between themselves and we are left with
\begin{align}
\hspace{-1.5em}	\dif_{\xi}\big\langle \prod_{i=1}^{n}\bar{\As}_{i}(x_{i})\prod_{j=1}^{N}K^{x^{\prime}_{\pi(j)}x_{j}}_{j}\big\rangle_{\xi}  = \sum_{m=1}^{n} \sum_{\{ \rho_{m} \}} (-1)^{m-1} \dif_{\xi}\big \langle \prod_{i=1}^{m} i \dsp_{\rho(i)} \prod_{j = m+1}^{n} \bar{\As}_{\rho(j)} \prod_{k=1}^{N}K^{x^{\prime}_{\pi(k)}x_{k}}_{k}\big \rangle_{\xi}\, ,
\label{eqDeltaXiAs2}
\end{align} 
where the sum over permutations sums all possible replacements of $m$ of the $\bar{A}$ with partial derivatives.  Indeed, the $n = 1$ case has been given in (\ref{eqDifEx}) and the general case is proven with strong induction as follows. 

We suppose the result holds for up to $n$ insertions and use (\ref{eqDeltaXiAs}) and (\ref{eqDerivs}) for the case of $n+1$ insertions. This leads straightforwardly to
\begin{align}
	\hspace{-1em}\dif_{\xi}\big\langle \prod_{i=1}^{n+1}\bar{\As}_{i}(x_{i})&\prod_{j=1}^{N}K^{x^{\prime}_{\pi(j)}x_{j}}_{j}\big\rangle_{\xi}  = \sum_{m=1}^{n+1}\sum_{\{\rho_{m}\}} (-1)^{m-1}\Big[ \dif_{\xi} \big \langle \prod_{i = 1}^{m}i \dsp_{\rho(i)} \prod_{j=m+1}^{n+1} \bar{\As}_{\rho(j)} \prod_{k = 1}^{K} K^{x^{\prime}_{\pi(k)}x_{k}}_{k} \big \rangle_{\xi} \nonumber \\
	\hspace{-1em}&- \sum_{p=1}^{m}\sum_{\{\sigma_{p}\}} \prod_{q = 1}^{p} (-1)^{p-1}i \dsp_{\rho(\sigma(q))} \dif_{\xi} \big \langle  \prod_{r = p+1}^{m} i \dsp_{\rho(\sigma(r))} \prod_{j = m+1}^{n+1} \bar{\As}_{\rho(j)} \prod_{k = 1}^{K} K^{x^{\prime}_{\pi(k)}x_{k}}_{k} \big \rangle_{\xi} \Big]\, .
\end{align}
The terms on the first line are precisely the result desired; it remains to show that the sum of terms coming from the second line cancel. To verify this, we consider the sum of all terms that contain $p = s$ derivatives outside of the variation $\diff_{\xi}$. The terms involving such derivatives are
\begin{equation}
	\sum_{m = s}^{n+1}\sum_{\{\rho_{m}\}} (-1)^{m+s-1} \sum_{\{\sigma_{s} \}} \prod_{q=1}^{s} i \dsp_{\rho(\sigma(q))} \dif_{\xi} \big \langle \prod_{r = s+1}^{m} i \dsp_{\rho(\sigma(r))}  \prod_{j=m+1}^{n+1}\bar{\As}_{\rho(j)} \prod_{k = 1}^{K} K^{x^{\prime}_{\pi(k)}x_{k}}_{k} \big \rangle_{\xi} \,.
\end{equation} 
Of these terms, when $m = s$ there are no derivatives in the expectation value, and there are $n$ or fewer factors of $\bar{A}$. The inductive hypothesis shows that this term,
\begin{equation}
 	-\sum_{\{\rho_{s}\}} \sum_{\{\sigma_{s} \}} \prod_{q=1}^{s} i \dsp_{\rho(\sigma(q))} \dif_{\xi} \big \langle \prod_{j=s+1}^{n+1}\bar{\As}_{\rho(j)} \prod_{k = 1}^{K} K^{x^{\prime}_{\pi(k)}x_{k}}_{k} \big \rangle_{\xi} \, ,
\end{equation}
cancels against the terms that have $1$ or more derivatives inside the brackets. This completes the proof. 

The immediate application is to derive the LKF transformation. It is clear that
\begin{equation}
	\dif_{\xi} \big \langle \prod_{i=1}^{n} \big[m + i\dsp_{i} - \As^{\gamma}_{i} - \bar{\As}_{i}\big] \prod_{k = 1}^{K} K^{x^{\prime}_{\pi(k)}x_{k}}_{k}[A^{\gamma} + \bar{A}] \big \rangle_{\xi} = 0
\end{equation}
since the signs and the derivatives in the variation of $\diff_{\xi}\big\langle \prod_{i=1}^{n}\bar{\As}_{i}(x_{i})\prod_{j=1}^{N}K^{x^{\prime}_{\pi(j)}x_{j}}_{j}\big\rangle_{\xi}$ given in (\ref{eqDeltaXiAs2}) cancel term by term against the variations with fewer $\bar{\As}$. To illustrate how the cancellation works, we work out the $n = 2$ case explicitly, organising the calculation according to powers of $[m - \As^{\gamma}]$ whose gauge variation is trivial:
\begin{align}
	\dif_{\xi}&\big\langle \big[m + i\dsp_{1} - \As^{\gamma}_{1} - \bar{\As}_{1}\big]\big[m + i\dsp_{2} - \As^{\gamma}_{2} - \bar{\As}_{2}\big] K^{x^{\prime}_{\pi(1)}x_{1}}_{1}K^{x^{\prime}_{\pi(2)}x_{2}}_{2} \big \rangle_{\xi} \nonumber\\
		=&\big[m - \As^{\gamma}_{1} \big]\big[m  - \As^{\gamma}_{2}\big] \dif_{\xi}\big \langle K^{x^{\prime}_{\pi(1)}x_{1}}_{1}K^{x^{\prime}_{\pi(2)}x_{2}}_{2} \big \rangle_{\xi} \\
		+& \big[m - \As^{\gamma}_{1} \big] \dif_{\xi}\big \langle \big[i\dsp_{2} - \bar{\As}_{2}\big] K^{x^{\prime}_{\pi(1)}x_{1}}_{1}  K^{x^{\prime}_{\pi(2)}x_{2}}_{2} \big \rangle_{\xi} \\ 
		+& \big[m - \As^{\gamma}_{2} \big] \dif_{\xi}\big \langle \big[i\dsp_{1} - \bar{\As}_{1}\big] K^{x^{\prime}_{\pi(1)}x_{1}}_{1} K^{x^{\prime}_{\pi(2)}x_{2}}_{2}\big \rangle_{\xi}		\\
		+& \dif_{\xi} \big \langle \big[i\dsp_{1} - \bar{\As}_{1}\big]\big[i\dsp_{2} - \bar{\As}_{2}\big] K^{x^{\prime}_{\pi(1)}x_{1}}_{1} K^{x^{\prime}_{\pi(2)}x_{2}}_{2}\big \rangle_{\xi} \, .
\end{align}
Now we already know that $\diff_{\xi}\big \langle K^{x^{\prime}_{\pi(1)}x_{1}}_{1}[A^{\gamma}+\bar{A}] K^{x^{\prime}_{\pi(2)}x_{2}}_{2}[A^{\gamma} + \bar{A}] \big \rangle_{\xi} = 0$, and in fact our $n = 1$ case from above shows that the second two lines also vanish.  For the last line we begin with the term involving $\bar{\As}_{1} \bar{\As}_{2}$:
\begin{align}
	\dif_{\xi} \big \langle  \bar{\As}_{1} \bar{\As}_{2} K^{x^{\prime}_{\pi(1)}x_{1}}_{1} K^{x^{\prime}_{\pi(2)}x_{2}}_{2}\big \rangle_{\xi}  = &-\dif_{\xi}\big \langle i \dsp_{1} i \dsp_{2} K^{x^{\prime}_{\pi(1)}x_{1}}_{1}  K^{x^{\prime}_{\pi(2)}x_{2}}_{2}\big \rangle_{\xi} \nonumber \\
	&+ \dif_{\xi}\langle i\dsp_{1} \bar{\As}_{2} K^{x^{\prime}_{\pi(1)}x_{1}}_{1} K^{x^{\prime}_{\pi(2)}x_{2}}_{2} \big \rangle_{\xi} \nonumber \\
	&+\dif_{\xi}\langle \bar{\As}_{1}i\dsp_{2}K^{x^{\prime}_{\pi(1)}x_{1}}_{1}  K^{x^{\prime}_{\pi(2)}x_{2}}_{2}\big \rangle_{\xi}
	\label{eqAA2}
\end{align}
which cancel the other terms that arise in the final line. The variations therefore sum to zero.

For completeness we also exhibit how the result (\ref{eqDeltaXiAs2}), used in this illustration, arises in this simple case. We can use (\ref{eqDeltaXiAs}) and the commutator to write
\begin{align}
	\dif_{\xi} \big \langle  \bar{\As}_{1} \bar{\As}_{2} K^{x^{\prime}_{\pi(1)}x_{1}}_{1} K^{x^{\prime}_{\pi(1)}x_{1}}_{k} \big \rangle_{\xi} &= \Big[ \big \langle \bar{\As}_{2} K^{x^{\prime}_{\pi(1)}x_{1}}_{1} K^{x^{\prime}_{\pi(2)}x_{2}}_{2}\big \rangle_{\xi} i \dsp_{1} \nonumber \\
	&+  \big \langle \bar{\As}_{1} K^{x^{\prime}_{\pi(1)}x_{1}}_{1}K^{x^{\prime}_{\pi(2)}x_{2}}_{2} \big \rangle_{\xi} i \dsp_{2} +  \big \langle  K^{x^{\prime}_{\pi(1)}x_{1}}_{1}K^{x^{\prime}_{\pi(2)}x_{2}}_{2}\big \rangle_{\xi} i\dsp_{1} i \dsp_{2}\Big] \e^{-\Delta_{\xi}S}\\
	&= \dif_{\xi}\big \langle i\dsp_{1} \bar{\As}_{2}   K^{x^{\prime}_{\pi(1)}x_{1}}_{1}  K^{x^{\prime}_{\pi(2)}x_{2}}_{2}\big \rangle_{\xi} - i\dsp_{1}\dif_{\xi}\big\langle   \bar{\As}_{2} K^{x^{\prime}_{\pi(1)}x_{1}}_{1} K^{x^{\prime}_{\pi(2)}x_{2}}_{2} \big \rangle_{\xi} \nonumber \\
	&+  \dif_{\xi}\big \langle i\dsp_{2} \bar{\As}_{1}   K^{x^{\prime}_{\pi(1)}x_{1}}_{1} K^{x^{\prime}_{\pi(2)}x_{2}}_{2}\big \rangle_{\xi} - i\dsp_{2}\dif_{\xi}\big\langle   \bar{\As}_{1} K^{x^{\prime}_{\pi(1)}x_{1}}_{1} K^{x^{\prime}_{\pi(2)}x_{2}}_{2}\big \rangle_{\xi}  \nonumber \\
	&- \dif_{\xi}\big \langle i \dsp_{1} i \dsp_{2} K^{x^{\prime}_{\pi(1)}x_{1}}_{1} K^{x^{\prime}_{\pi(2)}x_{2}}_{2}\big \rangle_{\xi} + i \dsp_{1}\dif_{\xi}\big \langle i\dsp_{2}K^{x^{\prime}_{\pi(1)}x_{1}}_{1}  K^{x^{\prime}_{\pi(2)}x_{2}}_{2} \big \rangle_{\xi} \nonumber \\
	&+ i \dsp_{2}\dif_{\xi}\langle i \dsp_{1}K^{x^{\prime}_{\pi(1)}x_{1}}_{1} K^{x^{\prime}_{\pi(2)}x_{2}}_{2} \big \rangle_{\xi} -  i\dsp_{1}i\dsp_{2} \dif_{\xi}\big \langle K^{x^{\prime}_{\pi(1)}x_{1}}_{1}  K^{x^{\prime}_{\pi(2)}x_{2}}_{2}\big \rangle_{\xi} \,.
\end{align}
The last two lines can be verified by direct calculation (the last term vanishes, of course) to verify (\ref{eqDerivs}) and (\ref{eqAA2}). The $n = 1$ case can again be used to cancel all but the three terms that make up (\ref{eqDeltaXiAs2}). In the course of working out this example in detail we have also verified the alternating signs that enter the equations (\ref{eqDerivs}) and (\ref{eqDeltaXiAs2}).

\bigskip

\bibliography{bibLKFT}
\end{document}